\documentclass{aa}  

\usepackage{graphicx}
\usepackage{txfonts}
\usepackage{hyperref}
\newcommand{\coii}{$\rm CO1.58$}
\newcommand{\coiii}{$\rm CO1.60$}
\newcommand{\coiv}{$\rm CO1.64$}
\newcommand{\cov}{$\rm CO1.66$}
\newcommand{\covi}{$\rm CO1.68$}
\newcommand{\covii}{$\rm CO2.30$}
\newcommand{\coviii}{$\rm CO2.32$}
\newcommand{\coix}{$\rm CO2.35$}
\newcommand{\zh}{$\rm [Z/H]$}
\newcommand{\mh}{$\rm [M/H]$}
\newcommand{\cfe}{$\rm [C/Fe]$}
\newcommand{\mgfe}{$\rm [Mg/Fe]$}
\newcommand{\ofe}{$\rm [O/Fe]$}
\newcommand{\xfe}{$\rm [X/Fe]$}
\newcommand{\nafe}{$\rm [Na/Fe]$}
\newcommand{\nfe}{$\rm [N/Fe]$}
\newcommand{\cafe}{$\rm [Ca/Fe]$}
\newcommand{\tife}{$\rm [Ti/Fe]$}
\newcommand{\sife}{$\rm [Si/Fe]$}
\newcommand{\tioi}{$\rm TiO1$}
\newcommand{\tioiir}{$\rm TiO2_{SDSS}$}
\newcommand{\fef}{$\rm Fe5015$}
\newcommand{\cnt}{$\rm CN2$}
\newcommand{\cno}{$\rm CN1$}
\newcommand{\caf}{$\rm Ca4227$}
\newcommand{\cat}{$\rm CaT$}
\newcommand{\afe}{$\rm [\alpha/Fe]$}
\newcommand{\kms}{\,km\,s$^{-1}$}
\newcommand{\VROT}{$\rm V_{rot}$}
\newcommand{\SIG}{$\rm \sigma$}

\newcommand{\gammab}{$\rm \Gamma_b$}
\newcommand{\mgfep}{$\rm [MgFe]'$}

\begin{document}

\titlerunning{CO radial gradients in the bulge of M31}
\authorrunning{F. La Barbera et al.}

   \title{Radial gradients of CO absorptions and abundance ratios in the bulge of M31}

   \subtitle{}

   \author{F. La Barbera~\inst{1} \and A. Vazdekis~\inst{2, 3} \and A. Pasquali~\inst{4} \and J. Heidt~\inst{5} \and A. Gargiulo~\inst{6} \and E. Eftekhari~\inst{2, 3}
          }

   \institute{INAF-Osservatorio Astronomico di Capodimonte, sal. Moiariello
16, Napoli, 80131, Italy\\
              \email{francesco.labarbera@inaf.it}
         \and
             Instituto de Astrof\'\i sica de Canarias, E-38205 La Laguna, Tenerife, Spain
         \and
         Departamento de Astrof\'\i sica, Universidad de La Laguna (ULL), E-38206  La Laguna, Tenerife, Spain
         \and
         Astronomisches Rechen-Institut, Zentrum f\"ur Astronomie, Universit\"at Heidelberg, M\"onchhofstr. 12-14, D-69120 Heidelberg, Germany
         \and
         Landessternwarte, Zentrum f\"ur Astronomie der Universit\"at Heidelberg, K\"onigstuhl 12, 69117 Heidelberg, Germany
         \and
         INAF – Istituto di Astrofisica Spaziale e Fisica Cosmica Milano, Via A. Corti 12, 20133 Milano, Italy
             }

   \date{Received ; accepted}

 
   \abstract{
     We present new H- and K-band spectroscopy for the bulge of M31, taken with the LUCI spectrograph at the Large Binocular Telescope (LBT). 
We studied radial trends of CO absorption features (namely, \coii , \coiii, \coiv , \cov , \covi , \covii , \coviii , and \coix  
) in the bulge of M31, out to a galactocentric distance of $\sim 100$'' ($\sim 380$~pc). We find that most COs do not exhibit a strong radial gradient, despite the strong metallicity gradient inferred from the optical spectral range, except for \coiv , showing a steep increase in the center.
We compared the observed line strengths to predictions of different state-of-the-art  stellar population models, including an updated version of EMILES models, which also uses the extended IRTF spectral library. The observed COs are close to models' predictions, but in some models they turn out to be underestimated. We find that the lack of radial gradients is due to the combination of increasing CO strength with metallicity and C abundance, and decreasing CO strength with IMF slope and O abundance. We speculate that the steep gradient of \coiv\ might be due to Na overabundance.
Remarkably, we were able to fit, at the same time, optical indices and all the NIR COs (except for CO1.68), { leaving abundance ratios (i.e., \cfe, \ofe, and \mgfe ) as free-fitting parameters, imposing age and metallicity constraints from the optical spectral range}, with no significant contribution from intermediate-age populations ($\sim 1$~Gyr-old). For the majority of the bulge, we find \mgfe$\sim 0.15$~dex, \ofe\ larger than \mgfe\ (by $\sim 0.1$~dex), and C abundance consistent with that of Mg. In the central (few arcsec) region, we still find an enhancement of O and Mg, but significantly lower \cfe .  We find that the COs' line strengths of the bulge are significantly lower than those of massive galaxies, possibly because of a difference in carbon abundance, as well as, to some extent, total metallicity.
   }


   \maketitle
%
   \nolinenumbers
   
\section{Introduction}

The detailed chemical composition of a stellar system is a unique imprint of how it formed and evolved. Different chemical elements are released into the interstellar medium (ISM) by different types of stars, with different masses, and therefore over different timescales. The proportion of such stars is given by the stellar IMF, that is, the mass distribution of stars at birth in a stellar population. Chemical elements  are then incorporated into new generations of stars depending on the star formation efficiency of the system.  Therefore,  abundance ratios of different elements can be used as clocks of star formation, that is, to constrain star formation timescales, and to probe the overall star formation efficiency, as well as the stellar IMF.

The analysis of spectral absorption features -- sensitive to the presence of low-mass stars in a stellar population, such as the
Na I doublet feature at $\lambda \lambda$8183, 8195~\AA\ (\citealt{FaberFrench:80};  \citealt{SchiavonFeH:97}, hereafter NaI8200) -- has
shown that the IMF of massive early-type galaxies (ETGs) is not universal, with an excess of low-mass stars, that is, a bottom-heavy IMF, in the central regions of more massive galaxies  (see,
e.g., \citealt{vdC:10,Ferr:13,LB:13, Spiniello:2014}).
This technique relies on the comparison of observed line strengths with predictions of stellar population models. A crucial aspect is that IMF-sensitive features are also sensitive to chemical abundance ratios, and they are potentially affected by uncertainties on stellar population models~\citep{CvD12a}. In order to overcome these degeneracies, one has to analyze as many  features as possible, from different chemical species over a large wavelength range.
While in the optical spectral range different features do indeed provide consistent IMF constraints, the situation  is less clear in the Near-Infrared (NIR; $\lambda \gtrsim 1$~$\mu$m; see, e.g., ~\citealt{Alton:2017, Alton:2018, elham:2021a}). One major problem is to obtain accurate stellar population models in the NIR, from { empirical} infrared stellar spectra as good as those available in the optical, also covering the stellar parameters' space  as extensively as
possible (in particular, in the super-solar metallicity regime, and for hot stars). { Only recently has some progress been made in this direction, thanks to the advent of the extended IRTF~\citep{Villaume:2017} and XSL~\citep{Verro:2022a} empirical spectral libraries. We note that, in contrast to empirical libraries, theoretical stellar libraries are available for both a large range in wavelength and in stellar parameters~(see, e.g., \citealt{Allard:2012, Husser:2013}, and references therein); however, they are hampered by large uncertainties in the calculation
of model atmospheres, especially in the low-temperature regime, which is relevant in the NIR (see, e.g., \citealt{Lancon:2021}).}

In order to apply state-of-the-art stellar population models in the NIR spectral range, one has to benchmark model predictions against high-quality spectroscopic data. One such benchmark is provided by the bulge of M31. Due to its proximity, the bulge of Andromeda allows us to obtain high S/N spectroscopy, at a low observational cost, also in the NIR spectral range, and to apply the same techniques used to constrain the stellar population content of ETGs. Moreover, metallicities and abundance ratios in the bulge can be constrained through resolved  (giant) stars~\citep{SJ:2005}, and the star formation history can be constrained through the { color-magnitude diagram (CMD; ~\citealt{Dong:2018})}.
Most of the stars in the bulge are old and enhanced in alpha elements, as in massive galaxies~\citep{Saglia:2010}. Metallicity is around solar for most of the bulge, giving us the opportunity to study the effect of abundance ratios in a metallicity range where stellar population models provide safe predictions. Moreover, in the central (few arcsec) region,  the bulge exhibits super-solar metallicity, similar to massive ETGs~\citep{Saglia:2010, Saglia:2018}{ . This also allows} us to test the models in the super-solar metallicity regime that characterizes the central regions of massive galaxies.

The stellar IMF of the M31 bulge has been the subject of an intense debate over the years. Because of the strong NaI8200 feature, \citet{SpinTa:71} concluded in favor of a dwarf-dominated IMF, with a high mass-to-light ratio ($\rm M/L=44$).
However, evidences against a bottom-heavy IMF were found by~\citet{Whitford:77} and~\citet{Cohen:1978}, based on the FeH Wing-Ford band~\citep{SchiavonFeH:97}, as well as by~\citet{Faber:1972} and~\citet{Baldwin:1973a}, based on the first overtone of CO in the K band at $\sim$2.3~$\mu$m (hereafter CO2.3).
In particular, these latter studies found CO2.3 to be significantly stronger in the bulge than what was expected for a dwarf-enriched population, with this feature being more prominent in giant, rather than dwarf, stars.
Interestingly, along the same lines,~\citet{Frogel:1978} claimed that the K-band CO of ETGs was consistent with a giant-, rather than a dwarf-, dominated population. In particular, \citet{Frogel:1980} attributed the CO2.3 absorption in ETGs to the presence of low-temperature luminous stars, such as giant branch stars with super-solar metallicity, and/or asymptotic giant-branch (AGB) stars. Since that time, many works have used CO line strengths to infer the presence of young stars (red giants and super-giants) in ETGs~\citep{MJ:1996, JM:1999, MJ:2000, MarmolQueralto:2009, Maraston:2005, Zibetti:2013}.

{ Nowadays, based on dynamical arguments, we know that the bulge of M31 cannot have a very bottom-heavy IMF.
The low velocity dispersion of the bulge, $\sim 150$~\kms\ (see \citealt{Saglia:2010}; \citealt{paperI}, hereafter Paper I), strongly constrains the dynamical mass-to-light ratio ($\rm M/L_D$), setting an upper limit to the stellar $\rm M/L$ (i.e., the IMF normalization).~\citet{Saglia:2010} did indeed  find that the $\rm M/L_D$ is consistent with the stellar $\rm M/L$ estimated for a Kroupa IMF.
In fitting IMF-sensitive features with predictions of state-of-the-art stellar population models with} the same technique applied to the spectra of massive ETGs, recent studies have found that the stellar IMF of the bulge is mostly consistent with a Kroupa-like distribution, with a mildly bottom-heavy population confined only to the central region, { giving a negligible contribution to the total stellar $\rm M/L$ of the bulge} ( \citealt{CvD12b}; Paper I). Since no attempt has been made so far to { fit a} homogeneous set of optical and NIR data, the question remains of whether the results obtained from the optical spectral range are consistent with constraints from the NIR spectral range and, in particular, from giant-sensitive CO features.

{ Given the progress } in the field of stellar population synthesis as well as the far better quality of NIR spectroscopy currently, it is now the time to revisit the questions raised from the early studies in the 1970s.
Recently, \citet[hereafter ELV22a]{elham:2022a} performed a systematic investigation of NIR CO absorptions in massive ETGs, extending previous works targeting the K-band CO2.3 feature. Several CO absorptions dominate the H-band spectral region of old stellar populations, as well as the red end of the K band. ELV22 found that all the NIR COs are systematically under-predicted by state-of-the-art, scaled-solar, stellar population models. The discrepancy is apparently even more significant when accounting for the fact that massive ETGs have a bottom-heavy IMF, as inferred from the optical spectral range.
However, \citet{Baldwin:2018} also found that the K-band CO is under-predicted by the models for a sample of low-mass ETGs, which are not expected to have a bottom-heavy IMF. Therefore, it is not clear if the discrepancy arises from some issue in the models, or some stellar population property of massive galaxies.
To investigate this issue, \citet[hereafter ELV22b]{elham:2022b} studied CO features in the spectra of the massive, relic, galaxy NGC\,1277 -- for which the presence of intermediate-age/young populations is extremely unlikely -- finding that the CO lines in this galaxy are as strong as in other massive ETGs. Therefore, ELV22a,b suggest that the strength of CO features in massive galaxies is not due to intermediate-age populations, but to some other effect, such as, in particular, carbon overabundance.

{ The present paper follows the stellar population analysis of the bulge of M31 presented in Paper I, which relied on high-S/N optical spectroscopy along the bulge major axis,  acquired with the Gran Telescopio CANARIAS (GTC) from the U band through the I band. In Paper I, we studied radial gradients of several spectral indices (including TiOs, Ca, and Na features) in the bulge of M31, finding that outside a galacto-centric distance of $\sim 10$'', the stellar IMF of the bulge is similar to a Milky-Way-like distribution, while at small galactocentric distances, an IMF radial gradient is detected, with a mildly bottom-heavy IMF in the few inner arcsec. Fitting Na absorptions in the optical, we also found that Na abundance ratios vary from $\sim$0.4~dex for most of the bulge, up to $\sim$0.6~dex in the innermost region.
} 

In the present paper, we aim to investigate the origin of CO absorptions in unresolved stellar populations, by studying NIR CO lines in the bulge of M31.
To this effect, we relied on new H- and K-band spectroscopy taken at the Large Binocular Telescope (LBT), with the  LBT Utility Camera in the Infrared (LUCI) instruments. 
{ We point out that the analysis of NIR absorption features in galaxies is still in its infancy. The contribution of different chemical species to NIR lines is still not completely understood (see e.g.,~\citealt{RV:17}). Moreover, NIR light is dominated by cool evolved stars for which the computation of synthetic stellar spectra, including the effect of varying abundance ratios, is uncertain. NIR stellar libraries that are used to construct stellar population models are also far less complete than those available in the optical. Therefore, for the present work, we treated the constraints on age, metallicity, and IMF from the optical (Paper I) as a reference, and scrutinized if the optical solution is consistent or if it shows some tension with the NIR COs.}
To analyze the H- and K-band data, we adopted an updated version of EMILES stellar population models~\citep{Vazdekis:2016}, and we performed an extensive comparison with predictions from other state-of-the-art stellar population  models, namely, CvD18~\citep{CvD18} and XSL~\citep{Verro:2022b} models. Our main goal is to investigate whether the mismatch between observed and model CO line strengths for massive galaxies results from some issue in the models that should also apply to the bulge of M31, or some other effect. Moreover, we want to compare model predictions with observed line strengths in the center of the bulge, where metallicity is super-solar, as in massive ETGs, and the IMF is mildly bottom-heavy. Our analysis takes the effect of abundance ratios into account -- such as carbon, oxygen, and magnesium -- on CO lines, presenting a first, joint, determination of such abundances for the bulge of M31. In a forthcoming paper, we will present a detailed analysis of other abundance ratios in the bulge.

The layout of the paper is as follows. 
In Sec.~\ref{sec:data}, we describe the new LBT data for the M31 bulge, as well as our procedure to extract radially binned spectra, and the data of massive galaxies. Sec.~\ref{sec:models} describes the stellar population models used in the analysis. The spectral indices adopted in the present work are described in Sec.~\ref{sec:indices}, together with our method to fit the indices and constrain the abundance ratios. Sec.~\ref{sec:results} shows our main results, that is, the radial profiles of CO indices in the bulge (Sec.~\ref{sec:COrad}), a comparison of different model predictions of CO indices (Sec.~\ref{sec:comp_models}), and the results of fitting CO and optical lines with different models, including the effect of abundance ratios  (Sec.~\ref{sec:fits}). We discuss the results in Sec.~\ref{sec:discussion}, providing a summary in Sec.~\ref{sec:summary}.
As in Paper I, we adopted a distance of 785~kpc from the Milky Way to M31~\citep{McConnachie:2005}, implying a conversion scale of $\rm \sim 3.8$~pc/arcsec.


\section{Data}
\label{sec:data}

\subsection{Observations}
\label{sec:obs}
We obtained new long-slit spectroscopy in H and K band, along the major axis of the bulge of M31, using the LUCI instruments, LUCI1 and LUC2, on board of LBT.
The H-band spectra were acquired on October 2021 (LBT proposal IT-2021B-017; PI: F. La Barbera), with a total on-source exposure time of 2.4~hr. We performed three sequences of 16 individual exposures, each with an exposure time of 180s (DIT=180~s, NDIT=1). Small offsets (10''-wide) along the slit among different exposures were applied, in order to allow for an optimal removal of cosmic rays and detector artifacts. To perform sky subtraction, every two exposures we offset the telescope to an empty sky region, observing the  sky for half of the total on-source exposure time.
A telluric standard star was observed with the same setup of the bulge, before and after the science exposures. Moreover, to perform relative flux calibration, we observed a star of the IRTF spectral library~\citep{IRTFI}, right after the observations of M31.

The K-band data were acquired on January 2019 (run A) and October 2021 (run B). Both runs were performed as part of German LBTB time (PI: A. Pasquali). During run A, we observed with LUCI2 only, performing a single sequence of 24 exposures with an exposure time of 75s each (DIT=15, NDIT=5). During run B, we observed with both LUCI1 and LUCI2 in binocular mode, with a sequence of 24 exposures with an exposure time of 90s each (DIT=90, NDIT=1). In this way, the bulge was observed for a total exposure time of 1.7hr. To perform sky subtraction, we adopted a similar sky-offset strategy as for the H-band data.

For all runs, we used the 0.5''-wide slit, covering a spatial region of about $\pm$~100'' ($\sim 380$~parsec) around the photometric center of the bulge, with a spatial scale of 0.25''/pixel. The typical seeing FWHM of the observations was $\sim$1'' (FWHM). For both H and K, we observed with the G210 grating, using the H- and K-band filters, respectively. Combining all the available runs in each band, this setup resulted into a wavelength range of 1.54--1.74~$\mu$m (2.015--2.374~$\mu$m) for H (K) band, and a uniform instrumental resolution of $\sigma \! \sim \! 35$~\kms. It is important to note that we adopted this setup, with a resolution significantly lower than the typical velocity dispersion of the bulge ($\sim 150$~\kms), in order to perform an accurate correction of sky emission and telluric lines.

The NIR data complement high-S/N optical spectroscopy along the major axis of the bulge, taken with the OSIRIS  (Optical System for Imaging and low-Intermediate-Resolution Integrated Spectroscopy) instrument at GTC, with the U-, V-, R-, and I-band R2500 grisms. The optical spectra were acquired with a slit of width 0.4'', similar to that of the LUCI slit, with similar spectral resolution ($\sigma \sim 38$~\kms ) and spatial scale ($0.254$''/pixel) as the LUCI data, out to a galactocentric distance of $\sim 200$~arcsec (see Paper I).

\subsection{Data reduction}
\label{sec:red}
The data were reduced using dedicated FORTRAN and IRAF routines developed by the authors. All frames were dark corrected, trimmed, and flat-fielded using dome-flat frames. For what concerns wavelength calibration, we found that the position of sky lines changed significantly from one frame to the other, even for short exposures (e.g., those of the standard stars). { This is due to the imperfect correction of the instrument internal flexures, resulting
in a shift of the entire spectrum on the detector~\citep{Heidt:2018}. To account for this effect, } we first computed an initial wavelength solution, and then, for each frame, we adapted this solution through a third-order spline, to match the position of bright, isolated, sky lines.
For the H-band data, the initial wavelength solution was computed from one of the sky frames, while in the K band, where sky lines at the red end are too faint to be used, we relied on  Ar+Ne+Xe arc-lamp frames.
The typical accuracy of wavelength calibration turned out to be $\sim 5$~\kms\ (rms) in both bands.
The flux calibration of the K band was performed using a telluric standard star, assuming a black-body spectrum of suitable temperature (given the spectral type of the star), while in the H band, we used the spectrophotometric standard star Feige110, observed with the same instrumental setup as for the science exposures. Sky subtraction was performed by correcting each M31 frame with the sky exposures observed before and after the given frame. Sky lines were rescaled to match those in the M31 frame by using the software {\sc SKYCORR}~\citep{Noll:2014}, with the same procedure described in Paper I for the optical spectroscopy of M31. To perform telluric correction, we used a telluric standard star observed before and after the M31 frames. We also tried to use the software {\sc MOLECFIT}~\citep{Smette:2015, Kausch:2015}, but this method provided worst residuals compared to the telluric standards.
For each band, the sky-corrected frames of M31 were rectified and combined. At each reduction step, and for each science frame, a variance map was produced and updated during the reduction process to account for different sources of uncertainty. Variance maps were finally combined in the same way as the science data. 

\subsection{Radially binned spectra}
\label{subsec:binning}
For both the H and K band, we extracted radially binned spectra along both sides of the slit, with a similar procedure as for the optical OSIRIS data. The central bin was set to be 1.5$''$-wide around the photometric center of the galaxy, 
while for the other bins we adopted a minimum width of $0.75''$ ($\sim 3$ pixels), increasing it adaptively outward, to ensure a minimum signal-to-noise ratio, $\rm S/N_{min}$~\footnote{For each band, the S/N is computed per \AA, as the average value over the corresponding wavelength range.}. First, in order to derive the kinematics of the bulge, i.e. rotation velocity \VROT , and velocity dispersion, $\sigma$, we adopted $\rm S/N_{min}=70$. We preferred to adopt a conservative, larger, threshold than in the optical ($\rm S/N_{min}=25$; see Paper I), in order to minimize the effects of sky residuals, that have a prominent role when analyzing NIR spectra.
As detailed in App.~\ref{app:kin}, when using E-MILES stellar population models (see below), the kinematics of the bulge in H and K band is fully consistent with that derived in the optical.
As a second step, for each band, at each position along the spatial direction of the slit, the rotation velocity profile was interpolated and used to correct the two-dimensional spectrum of M31 to the restframe. To perform the stellar population analysis (see below), a second set of radially binned spectra was extracted with $\rm S/N_{min}=300$ and $150$ in H and K band, respectively. These values of $\rm S/N_{min}$ were chosen to have the highest possible S/N ratio, without overly increasing the bin size along the spatial direction of the slit.

\begin{figure*}
\begin{center}
 \leavevmode
 \includegraphics[width=16cm]{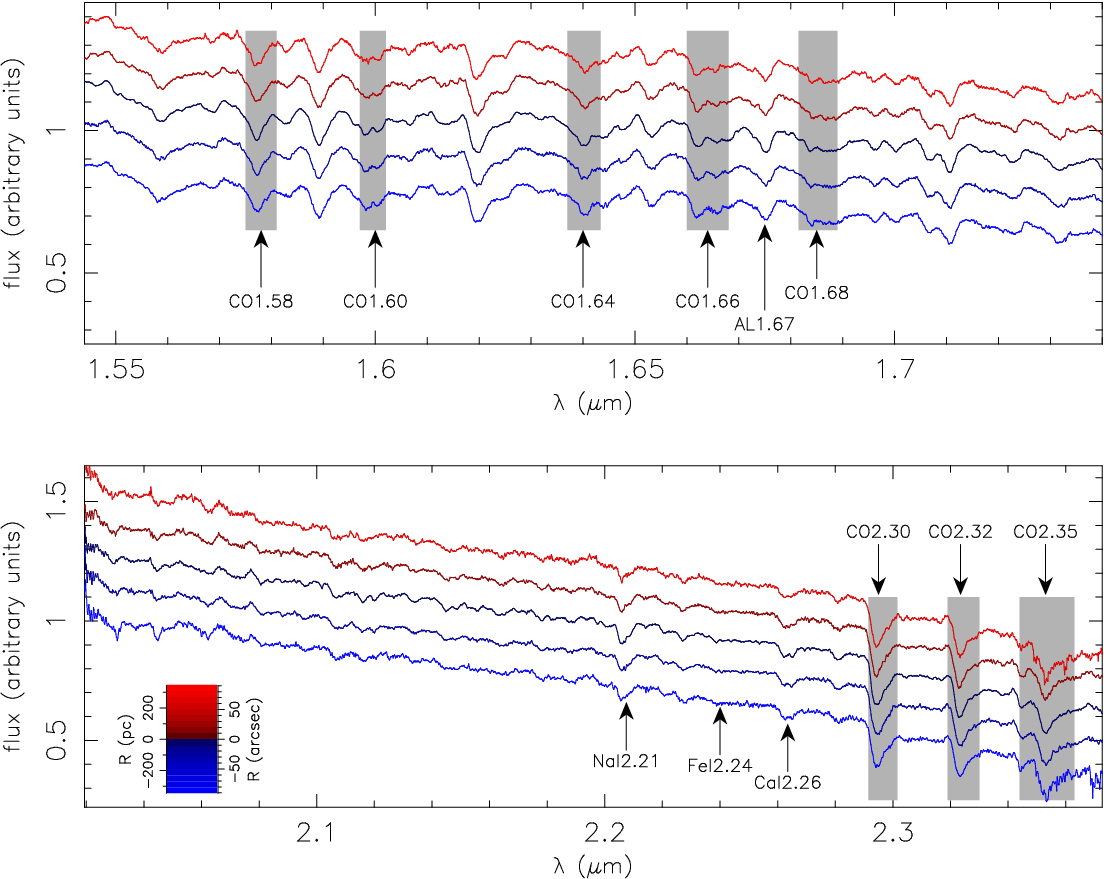}
\end{center}
 \caption{
H-(top) and K-(bottom) band radially binned spectra for the bulge of M31, observed with LUCI@LBT. In each panel, the main absorption features from~\citet[hereafter EVL21]{elham:2021a} are marked with a black arrow. Gray bands mark the sidebands of the CO absorption features analyzed in the present work, from H throughout K band (see the text).  Spectra with different colors correspond to different galactocentric distances. The distance to the center along the slit, $\rm R$,  is encoded with colors from blue (the southwest side of the slit) through red (northeast), as shown by the inset colored bar in the bottom panel.
 }
   \label{fig:M31spec}
\end{figure*}

Fig.~\ref{fig:M31spec} plots, as an example, some of the radially binned spectra of M31 in H (top) and K (bottom) band, showing the excellent quality of the LUCI data out to the largest galactocentric distances probed in the present work ($\rm R \sim 100$'', corresponding $\sim 380$~pc).

\subsection{Massive ETGs}
\label{subsec:XSGs}
We compare CO line strengths for the bulge of M31, with data for
seven massive ETGs, observed with the X-SHOOTER spectrograph at
ESO VLT (hereafter X-Shooter galaxies, XSG's; see~\citealt{LB:17, LB:19}). These objects were selected among the most massive ETGs of the SPIDER sample~\citep{SpiderI}, with a typical redshift of z=$0.05$, and a velocity dispersion range from 300 to 360~\kms.
Stacked spectra of the XSG's were constructed as described in ELV22a,
by combining high-S/N ($>$170~\AA$^{-1}$) spectra for individual galaxies,
extracted within a central aperture of radius 1.5'' ($\sim 1.4$~kpc).
As shown in LB19, the XSG's exhibit old stellar populations in their central regions (with an age of $\sim$11~Gyr), high metallicity (\zh\ between 0.15 and 0.4~dex), and \afe\ as high as 0.4~dex,  with
a significantly bottom-heavy IMF.
We note that only CO features bluer than CO2.30 could be measured on the XSG's spectra. Hence, our comparison with M31 excludes CO2.32 and CO2.35 (see below).

\section{Stellar population models}
\label{sec:models}
To analyze the spectra of M31, we rely on  EMILES~\citep{Vazdekis:2016}, \citet[hereafter CvD18]{CvD18}, and~\citet[hereafter XSL]{Verro:2022a, Verro:2022b} stellar population models~\footnote{We note that a comparison of EMILES and~\citet[{ hereafter M05}]{Maraston:2005} models  has been already presented in~ELV22a, and  therefore, it is not shown in the present paper.}.
EMILES  simple stellar population (SSP) models cover the spectral range from $0.168$ to $5 \mu$m, and are based on different empirical stellar libraries, namely the NGSL~\citep{Gregg:2006}, MILES~\citep{MILESI}, Indo-US~\citep{Valdes04},
CaT~\citep{CATI} and IRTF~\citep{IRTFI,IRTFII} stellar libraries (see also~\citealt{Vazdekis:12},~\citealt{RV:16}). 
{ In the present work, we use an updated version of EMILES models, that differs from the original one only in the NIR spectral range (from $8150$~\AA\ up to $\sim 2.5$~$\mu$m).  The computation of these models follows the same prescriptions/ingredients as for the original EMILES SSPs~\citep{RV:16}, with the only difference that in the NIR we add 205 stellar spectra from the extended IRTF~\citep{Villaume:2017} spectral library, to those of the original IRTF library (Vazdekis et al., 2024, in preparation).
We use the public available reddening-corrected { extended-IRTF} stellar spectra~\footnote{ https://irsa.ipac.caltech.edu/data/IRTF\_Extra/SpectralLibrary/}, having the same spectral resolution ($\sigma \sim 60$~\kms) as the IRTF spectra (see \citealt{RV:16} and references therein). Since all the { extended IRTF} stars are in common with the MILES stellar library, to construct the models, we use  MILES stellar parameters, namely $\rm \log g$, $\rm T_{eff}$, and $\rm [Fe/H]$, as in the original EMILES models, from the compilation of~\citet{Cenarro:2007}. In total, 380 (175 IRTF + 205 { extended IRTF}) stars are used in the updated models, providing a significantly better coverage in metallicity, especially in the subsolar metallicity range, with respect to the original models (see, e.g., figure~1 of \citealt{Villaume:2017}). For the present work, we are mostly interested to the improvement of the models in the solar and super-solar mettalicity regime, and for old (morer than a few { Gyr}) ages. At solar metallicity, all the updated models are safe, as the original EMILES SSPs (see~\citealt{Vazdekis:2016}). At higher metallicity, the quality parameter of the models (as defined, e.g., in eq.~4 of~\citealt{RV:16}) is around one, meaning that the models are approximately safe, improving~\footnote{ We note that for Teramo (Padova) models with \zh$=0.15$~dex ($0.22$~dex), the effective metallicity of the models is  close ($< 0.05$~dex) to the nominal one. On the other hand, for Teramo models with \zh$=0.26$~dex, the difference between the effective and nominal metallicity of the models is $\sim 0.07$~dex, meaning that even if the quality of the models is good, a further improvement of the stellar library coverage, at high \zh, is desirable.} significantly with respect to the original EMILES (see fig.~8 of~\citealt{RV:16}). 
In general, around solar metallicity, which is the regime of interest for the present work, 
the predictions of the update EMILES models are quite similar to those of the original models.
}

The EMILES models are computed for two sets of scaled-solar theoretical isochrones, namely
the ones of \citet{Padova00}
(Padova00; hereafter ``iP'') and those of \citet{Pietrinferni04} (BaSTI; hereafter ``iT''), the latter having lower temperatures  at the low-mass end~(see \citealt{Vazdekis:15}, and references therein).
The SSPs are computed for ages from $\sim 0.06$ to $\sim 17.8$\,Gyr ($0.03$ to $14$~Gyr), and metallicity, \zh, from $-2.2$ to $+0.22$~dex ($-1.7$ to $+0.26$~dex), for Padova00 (BaSTI)~\footnote{Note that BaSTI models are also computed for \zh$=0.4$ (see \citealt{Vazdekis:2016}). However, given the lower quality of these models, they are not used in the present analysis. Moreover, predictions of all models become unsafe for ages younger than $\sim 1$~Gyr in the NIR, unlike in the optical spectral range, due to the lack of hot stars in the IRTF library.} isochrones. As in Paper I, we use models computed for a ``bimodal'' IMF, i.e. single power-law distribution whose logarithmic slope, $\rm \Gamma_b$, is tapered at the low-mass end ($\lesssim 0.5 \, M_\odot$). Note that in this parametrization, increasing the high-mass end slope does also increase the dwarf-to-giant ratio in the IMF (implying a more bottom-heavy distribution) through its overall normalization. The bimodal IMF has been shown to provide mass-to-light ratios consistent with dynamical constraints~\citep{Lyubenova}. In Sec.~\ref{sec:discussion}, we also discuss the effect of changing the IMF shape. The EMILES models are computed for \gammab= \{$0.3, 0.5, 0.8, 1.0, 1.3, 1.5, 1.8, 2.0, 2.3, 2.5, 2.8, 3.0, 3.3, 3.5$\}. For $\rm \Gamma_b=1.3$, the bimodal IMF closely approximates the~\citet{Kroupa01} IMF.

Note that the stellar libraries used in the computation of EMILES, as well as XSL and CvD18 SSPs { (see below)} follow the abundance pattern of the Milky-Way, i.e. they are approximately scaled-solar at solar metallicity, and significantly $\alpha$--enhanced at metallicities roughly below $-0.2$~dex. Hence, EMILES, CvD18, and XSL,  SSPs should be considered as ``base'' (not scaled-solar) models. In the MILES spectral range (from $\lambda \! \sim \! 3500$ to $  \! \sim \! 7400$\AA), ~\citet{Vazdekis:15} also computed truly scaled-solar and $\alpha$--enhanced  SSPs, applying theoretical differential corrections for nonsolar \afe\ abundance ratios to the empirical SSPs. Although we do not use these models (hereafter $\alpha$--MILES) in the present work, we use constraints obtained with them from the optical spectral range (see Paper I).
Note also that, although not used in the present work, an updated version of $\alpha$--MILES (referred to as sMILES) models, has been recently published by~\citet{Knowles:2023}.

CvD18 models are an updated version of those of~\citet[hereafter CvD12]{CvD12a}. The models, covering the wavelength range 0.35--2.5~$\mu$m,  are based on the MILES stellar library in the optical, as well as the { extended IRTF}  stellar library in the NIR, the latter providing a better coverage in metallicity compared to the original IRTF (see above). The models are based on the MIST isochrones~\citep{Choi:2016, Dotter:2016}, and cover a range of ages (1--13.5~Gyr) and metallicities (from -1.5 to 0.2~dex). The SSPs are computed for different IMFs, using a three-segment parametrization, where one can change the slopes of the IMF between 0.1--0.5~$\rm M_\odot$ and 0.5--1~$\rm M_\odot$, while at higher masses the slope is fixed to that for a Salpeter distribution. The models also include a set of theoretical SSPs, computed for a Kroupa IMF, with varying abundance ratios for different elements, such as C, N, O~\footnote{CvD18 models vary O, Ne, and S in lock-step, referring to this variation as [$\alpha$s/Fe]. In the present paper, we assume that the main effect is that of O abundance, i.e. [O/Fe]=[$\alpha$s/Fe].}, Mg, Si, Ca, Ti, Fe, K, and Na, and others. Note that, in order to perform a meaningful comparison with EMILES models, and the data of M31, the CvD18 SSPs, whose wavelengths are originally in the vacuum, are converted to the air system~(see \citealt{Morton:1991}).

The XSL SSPs are based on the X-Shooter spectral library~\citep{Verro:2022a}, covering the wavelength range 0.3--2.5~$\mu$m, with an higher spectral resolution (R$\sim$10000) than EMILES and CvD18 models. The SSPs are computed for two sets of isochrones, namely Padova00 and PARSEC/COLIBRI (\citealt{Bressan:2012}; see also~\citealt{Verro:2022b} and references therein), and cover a range of ages, from 1 to 17.8~Gyr ($0.05$ to $15.8$~Gyr), and metallicities, from -1.7 to 0.2~dex (-2.2 to 0.2~dex), for Padova00 (PARSEC) isochrones. The models are computed for a Salpeter and a Kroupa IMF.

{ In Tab.~\ref{tab:SPmodels}, we provide  a coincise reference to the main features (such  as age and metallicity range,  IMFs, spectral  coverage,  and  resolution) of all SP models mentioned in the present work. }

\begin{table*}
        \centering
        \caption{
        {
        Summary of the features of stellar population   models quoted in the present work (see Sec.~\ref{sec:models}). Col.1 gives the name of the models. while col.~2 lists the isochrones used to construct the models. Cols.~3--5 provide the age and metallicity ranges, as well as the available IMFs.
        Cols.~6--7 list the spectral coverage and resolution of the models. All wavelengths are in the air system, but for CvD18 models, using the vacuum system.         }
}
        \label{tab:SPmodels}
\tiny
        \begin{tabular}{lcccccc}
        \hline
models & isochrones & age range & \zh\ range & IMFs & spectral coverage & resolution \\
(1) & (2) & (3) & (4) & (5) & (6) & (7) \\
\hline
EMILES iP\tablefootmark{a} & Padova00 & 0.0631--17.8~Gyr & -2.2--0.22~dex & unimodal, & 1680.2--49999.4~\AA\ & FWHM=3.0~\AA\ (<3060.8~\AA )\\
& & & & bimodal, && FWHM=5.0~\AA\ (3060.8-3541.4~\AA ) \\
& & & & Kroupa && FWHM=2.51~\AA\ (3541.4-8950.4~\AA ) \\
&&&&&& $\sigma$=60~\kms\ (>8950.4~\AA ) \\
EMILES iT\tablefootmark{a} & Teramo   & 0.03--14~Gyr &
-1.7--0.26~dex & // & // & // \\
$\alpha$--MILES\tablefootmark{b} & // & // & // & // & 3540.5-7409.6~\AA & FWHM=2.51~\AA\ \\
 sMILES\tablefootmark{b} & // & // & // & // & // & FWHM=2.51~\AA\ \\
 CvD18\tablefootmark{c} & MIST & 1--13.5~Gyr & -1.5--0.2~dex & two power-law IMF & 3501--24997.58~\AA\ & $\sigma$=100~\kms\ \\
 & & & & & & \\
 XSL-P00\tablefootmark{a} & Padova00 & 0.891--17.8~Gyr & -2.2--0.22~dex & Kroupa, Salpeter & 3500--24749.5~\AA\ & 11<$\sigma$<16~\kms\ \\
 XSL-PC\tablefootmark{a} & PARSEC   & 0.05--15.8~Gyr& -2.2--0.2~dex & // & // & // \\
 M05\tablefootmark{d} & Geneva, Teramo & $10^{-6}$--15~Gyr& -2.35--0.67~dex & Kroupa, Salpeter &  91~\AA -- 160~$\mu$m & $0.55 \le$FWHM~\tablefootmark{d}$ \le 11$~\AA \\  
\hline
        \end{tabular}
        \tablefoot{
        \tablefoottext{a}{Base models.}
        \tablefoottext{b}{Scaled-solar and $\alpha$-enhanced SSPs.}
        \tablefoottext{c}{Based models complemented with theoretical SSPs varying the abundance ratios of invidual elements.}
        \tablefoottext{c}{Based models; spectral resolution depends on stellar library, see~\citep{M11} for details.}
        }
\end{table*}


\section{Spectral indices}
\label{sec:indices}

\subsection{Index definitions}
\label{subsec:indices}
The LUCI spectra of M31 allow several CO absorption features to be studied in both H and K band. To this effect, we rely on the definition of Lick-like spectral indices from~\citet[hereafter EVL21]{elham:2021a}, who presented and fully characterized a large set of NIR spectral indices. In the H band, we analyze the CO1.58, CO1.60, CO1.64, CO1.66 and CO1.68 indices~\footnote{We did not include the CO1.56 index of EVL21 in our analysis, as this index is at the edge of the H-band spectral range, where LUCI spectra are severely affected by wavelength and flux calibration issues.}, while in the K band, we target CO2.30, CO2.32, and CO2.35. All these features, together with other prominent absorptions in each band are marked with black arrows in Fig.~\ref{fig:M31spec}. Central passband and pseudocontinua for CO indices are taken from table~1 of EVL21. In order to compare line strengths at different galactocentric distances with predictions of stellar population models, we first smoothed all LUCI spectra to the same sigma of $200$~\kms, i.e. the highest velocity dispersion of all radial bins, corresponding to the innermost radial bin of M31 (see App.~\ref{app:kin}), and then measured the spectral indices. { The smoothing was performed by convolving each spectrum with a Gaussian function, whose width was computed by subtracting in quadrature the velocity dispersion of the given spectrum and the instrumental sigma (35~\kms; see Sec.~\ref{sec:obs}) to a sigma of 200~\kms.} We note that SSP models used in our analysis were smoothed { to the same sigma} as the observed spectra.

In order to constrain the effect of abundance ratios on CO lines (Sec.~\ref{sec:fits}), we also analyze optical indices from the OSIRIS spectroscopy of the M31 bulge (see Paper I for details).  Indices are  measured after smoothing the optical spectra in the same way as the NIR ones. We use the optical indices Mgb, Fe5270, Fe5335, C4668, and Mg1, whose definitions are from \citet{Trager98}, as well as the composite metallicity indicator \mgfep$\rm =[ Mgb \cdot (0.72 Fe5270+0.28 Fe5335) ]^{1/2}$, defined by~\citet{TMB:03} to be insensitive to \mgfe\ abundance ratio (see also V15 and~\citealt{Knowles:2023}).

To { analyze the} optical and NIR spectra of M31 (Sec.~\ref{sec:fits}), we compute optical and NIR spectral indices for a representative ``inner'' and ``outer'' radial aperture. The inner aperture is defined by taking the average of line strengths within a given galactocentric distance, $\rm R_{in}$. 
{ Since we do not find strong radial variations for all COs but CO1.64 (see below), we set ~\footnote{ The use of a different $\rm R_{in}$ for CO1.64 and the other CO lines allows us to take into account the significant intrinsic scatter of CO radial trends in the computation of their average line strengths (see Sec.~\ref{sec:COrad}). However, we also verified that using the same aperture of $\rm R_{in}=3$'' for all COs would not change significantly our results.} $\rm R_{in}=10$'' for all COs but CO1.64. For the other indices, as well as CO1.64, we combine all radial bins within $3$''}. In the H and K bands, the outer aperture is defined by taking the mean value of line strengths, and the corresponding standard deviation (used as 1-sigma uncertainty), for all radial bins with absolute galactocentric distance between $35$'' and $100$'', corresponding to an average distance~\footnote{Note that outside a galactocentric distance of $\sim 30$'', both the optical and NIR radial profiles of spectral indices do not show any significant radial gradient, justifying the procedure to define an average outer aperture between $35$'' and $100$''.} from the center of $\rm R_{out}\sim 60$''. This procedure allows us to include in the uncertainties also the contribution from the intrinsic scatter of observed NIR line strengths (see Sec.~\ref{sec:COrad}). In the optical, the line strengths for the outer aperture are obtained by interpolating the observed line strengths' radial profiles, and their error bars (see figure~2 of Paper I), to a distance of $\rm R_{out}$.

\begin{figure*}
\begin{center}
 \leavevmode
 \includegraphics[width=16cm]{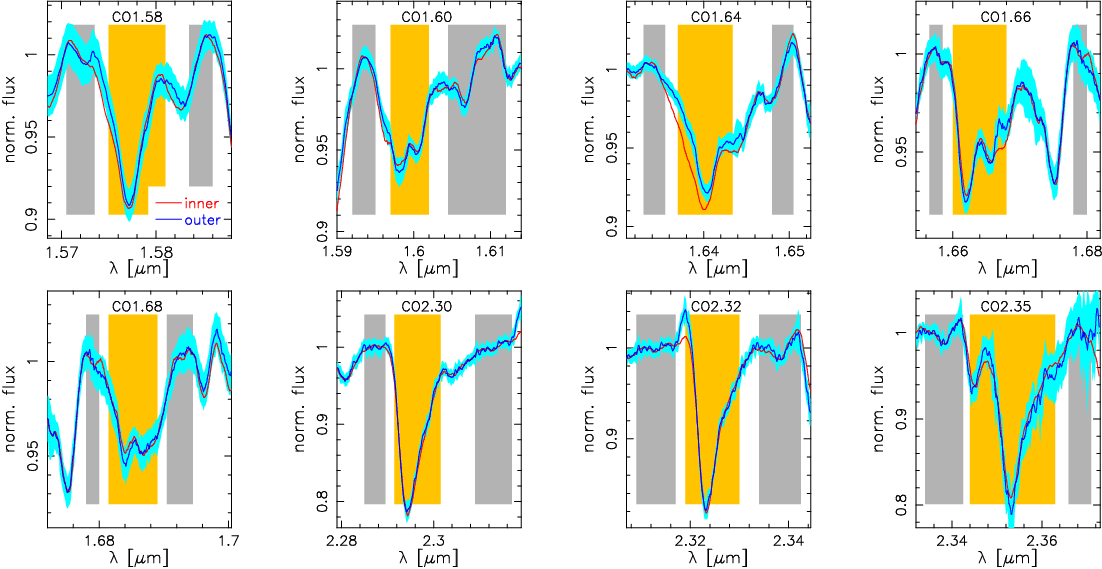}
\end{center}
 \caption{ Comparison of the ``inner'' (red) and ``outer'' (blue) spectra of M31 (see the legend in the top--left panel), in the spectral regions of the NIR CO features (see the text). Each panel corresponds to a different CO index, with wavelengths increasing from left to right, and top to bottom. The central passbands (pseudocontinuua) of the COs are marked with shaded orange (gray) regions.  For a given index, each spectrum is normalized by a line passing through the pseudocontinua regions. For illustrative purposes, we show  the (1-sigma) uncertainties on the outer spectrum only, plotted as a shaded cyan region.
 }
   \label{fig:spec_in_out}
\end{figure*}

{ Fig.~\ref{fig:spec_in_out} compares the inner and outer spectra of the M31 bulge, obtained by combining the available spectra in the same way as described above, for the line strengths. Note that the inner and outer spectra are fully consistent within the error bars, except for CO1.64, showing a stronger absorption in the inner aperture. We come back to discuss this point in Sec.~\ref{sec:results}. For Co2.32, an increase in flux at $\lambda \sim 2.3 \mu$m is seen for the outer aperture, likely due to an imperfect telluric correction. Since this flux excess is located between the blue pseudocontinuum and the central passband of the index, it does not affect significantly the measured line strengths.
}

\begin{table}
        \centering
        \caption{Scatter due to residual (excluding Mg, O, and C) abundance ratios, $\rm \sigma_{mod}$, on each spectral index considered in the present work. All values are in units of \AA, but that for Mg1 which is given in mag.}
        \label{tab:sigma_mod}
        \begin{tabular}{lc} 
      \hline
                index & $\rm \sigma_{mod}$ \\
 \hline
$\rm [MgFe]'$ & 0.04   \\
Mgb5177       & 0.055  \\
Fe5270        & 0.065  \\
Fe5335        & 0.060  \\
C4668         &  0.16  \\
Mg1           &  0.005 \\
CO1.58        &  0.05  \\
CO1.60        &  0.02  \\
CO1.64        &  0.07  \\
CO1.66        &  0.05  \\
CO1.68        &  0.06  \\
CO2.30        &  0.09  \\
CO2.32        &  0.50  \\
CO2.35        &  0.20  \\
 \hline
        \end{tabular}
\end{table}

\subsection{Index fitting}
\label{subsec:fit_indices}
In order to perform a quantitative analysis of CO absorption features, we fit line strengths with SSP model predictions.  We adopt an approach similar to that described in Paper I (see references therein), minimizing the expression:
\begin{eqnarray}
\rm \chi^2(Age, {\rm [Z/H]}, \Gamma_b, [X/Fe]_j) =
  \rm \sum_i \left[ \frac{E_{obs,i}- E_{mod,i} }{\sigma_i} \right]^2 + \nonumber \\
  \rm + \frac{(Age - Age_{opt})^2}{\sigma_{Age_{opt}}^2} + \frac{([Z/H] - [Z/H]_{opt})^2}{\sigma_{[Z/H]_{opt}}^2},
\label{eq:method}
  \end{eqnarray}
where the  index $\rm i$ runs over a selected  set of spectral features (see below); $\rm E_{obs,i}$ and $\sigma_i$ are observed
line strengths and their uncertainties; $\rm E_{mod,i}$ are predictions of line strengths from SSP models;
\xfe's are the elemental abundance ratios we aim to constrain; the second and { third terms} on the right side of Eq.~\ref{eq:method} account for  constraints on age and metallicity from the optical spectral range (see below). Note that according to CvD18 models, the main contributors to CO line strengths are, as expected,  [C/Fe] and [O/Fe], with some smaller contribution from [Mg/Fe], and minor contributions from other elements.
Therefore, in the present work, we mainly focus on \cfe, \ofe, and \mgfe~\footnote{Including \mgfe\ does also allow us to compare the abundance of O to that of Mg, which is traditionally used to estimate the abundance of alpha elements in unresolved stellar populations.}, treating the effect of other elemental abundances as an extra error budget on each index.
To this effect, the term $\rm \sigma_i$ in Eq.~\ref{eq:method} is obtained by summing up in quadrature the actual measurement error on the i-th index with an intrinsic error term, $\rm \sigma_{mod,i}$, given by the rms of the differences between line strengths of CvD18 models~\footnote{
We considered models with old age (13~Gyr) and solar metallicity (\zh $0$). However, we verified that none of our results is changing by selecting models with younger age (9~Gyr), or higher metallicity (\zh $=0.2$~dex).
} with enhanced Ca, Ti, Si, Na, N, and K, and that of the CvD18 solar-scale model.  The terms $\rm \sigma_{mod,i}$ provide an estimate of the contribution from residual abundance ratios (besides C, O, and Mg), to each index, and are reported~\footnote{For CO2.32, the actual value of $\rm \sigma_{mod}$ is $0.1$~\AA . However, this index turns out to decrease significantly with \nafe , with $\rm \delta(CO2.32)=-0.5$~\AA\ for $\rm \delta[Na/Fe]=+0.6$~dex. Since the bulge of M31 is significantly enhanced in \nafe (between 0.4 and 0.6~dex, see Paper I), the effect \nafe\ on CO2.32 is significant, justifying the choice to set $\rm \sigma_{mod}=0.5$~dex. Note this choice is not affecting any of our conclusions, as discussed in Sec.~\ref{sec:fits_abundances}.} in Tab.~\ref{tab:sigma_mod}. The effect of residual abundance ratios on our results is further discussed in Sec.~\ref{sec:fits}. Uncertainties on best-fitting parameters, namely Age, ${\rm [Z/H]}$, and \xfe's, are obtained from $\rm N=1000$ bootstrap iterations, where the fitting is repeated after shifting observed line strengths according to their uncertainties. 

\begin{table*}
        \centering
        \caption{Reduced chi-square and abundance ratios obtained by fitting line strengths for the ``inner'' ($R=0$'') and ``outer'' ($R= 60$'') reference radial bins of the M31 bulge, for different models and sets of indices (see the text).}
        \label{tab:fits_ews}
        \begin{tabular}{lcccccccccc} 
          \hline
         & & \multicolumn{4}{c}{INNER} & & \multicolumn{4}{c}{OUTER} \\
          \cline{3-6} \cline{8-11} \\
         fitting &  & $\chi^2_\nu $ & \cfe & \mgfe & \ofe &  & $\chi^2_\nu $ & \cfe & \mgfe & \ofe \\
          \hline
optical       & &   0.30 & $  0.04 \pm   0.02$ &$  0.14 \pm   0.02$ &$  0.02 \pm   0.08$ & &   0.30 & $  0.15 \pm   0.02$ &$  0.16 \pm   0.02$ &$  0.27 \pm   0.06$  \\
EMILES iT     & &   1.07 & $  0.03 \pm   0.01$ &$  0.15 \pm   0.02$ &$  0.05 \pm   0.05$ & &   1.87 & $  0.14 \pm   0.02$ &$  0.16 \pm   0.02$ &$  0.38 \pm   0.07$  \\
EMILES iP     & &   0.76 & $  0.04 \pm   0.02$ &$  0.12 \pm   0.02$ &$  0.07 \pm   0.06$ & &   1.11 & $  0.14 \pm   0.02$ &$  0.14 \pm   0.02$ &$  0.26 \pm   0.04$  \\
CvD18         & &   1.51 & $  0.06 \pm   0.02$ &$  0.08 \pm   0.02$ &$  0.01 \pm   0.08$ & &   1.23 & $  0.17 \pm   0.02$ &$  0.14 \pm   0.02$ &$  0.27 \pm   0.04$  \\
EMILES iT-IMF & &   1.94 & $  0.07 \pm   0.01$ &$  0.19 \pm   0.02$ &$  0.09 \pm   0.05$ & &   1.34 & $  0.16 \pm   0.02$ &$  0.15 \pm   0.02$ &$  0.37 \pm   0.08$  \\
\hline
  \end{tabular}
\end{table*}

{ In order to derive the best-fitting parameters in Eq.~\ref{eq:method}, namely Age, \zh, and \xfe\ (as the IMF slope, \gammab , is kept fixed in the present work; see below), for a given set of SSP models, we perform a linear interpolation of the models' line strengths  over a grid in age and metallicity, with steps of $0.1$~Gyr and $0.015$~dex, respectively. We consider only models with ages older than 1~Gyr, and metallicity $\rm [Z/H]>-0.5$~dex, as other values are  not relevant for the  bulge of M31. At each point in the grid, we write 
$\rm E_{mod,i} =  E_{mod,i,0} + \sum_X \Delta_{i,X} \cdot [X/Fe]$, 
where $\rm E_{mod,i,0}$ are predictions for solar-scale models (either EMILES, or CvD18, see below), and 
$\Delta_{i,X}$ is the sensitivity of the i-th spectral index to \xfe\ (see, e.g., ~\citealt{LB:15} for details), computed with CvD18 SSP models. Setting 
$\partial{\chi^2}/\partial{[X/Fe]}=0$, Eq.~\ref{eq:method} turns into a system of linear equations that is solved numerically, through a matrix inversion, providing the \xfe's, for each age and metallicity in the input grid. Final estimates of Age, \zh, and \xfe's are then obtained through a full exploration of the input grid, taking the values that minimize the $\chi^2$ (Eq.~\ref{eq:method}). 
}

{ To analyze the } CO line strengths (Sec.~\ref{sec:fits}),
we consider different cases (see Col.~1 of Tab.~\ref{tab:fits_ews}):

\begin{description}
 \item[{\it optical} - ] we fit only a selected set of  optical indices, i.e. \mgfep, Mgb, Fe5270, Fe5335, C4668, and Mg1. Since the quality of the fits does not depend significantly on the adopted model (i.e. EMILES Padova/Teramo or CvD models), we  average results from different models. Note that we consider the above set of optical indices as they are mainly sensitive to the abundance ratios we aim to constrain, i.e. [C/Fe], [Mg/Fe], and [O/Fe] (see, e.g.,  figure~1 of~\citealt{JTM:2012}), with smaller contribution from other abundances (which is accounted for by the residual scatter terms in Tab.\ref{tab:sigma_mod}); { note that since the origin of optical indices is better understood than that of NIR spectral lines, we regard the solution obtained in the optical as a reference in our analysis, scrutizing to what extent it can match the NIR CO line strengths};
 \item[{\it EMILES iT} - ] we fit simultaneously the optical indices (see above) and the NIR COs. We exclude CO1.68 as this feature is far off all models' predictions (see Sec.~\ref{sec:comp_models}), and including it would only drive the $\chi^2$ of the fits to significantly higher values, without affecting our main conclusions. The IMF is fixed to a Kroupa-like distribution (i.e. $\rm \Gamma_b=1.3$);
 \item[{\it EMILES iP} - ] same as EMILES iT but for models based on Padova isochrones;
 \item[{\it CvD} - ] same as EMILES iT and EMILES iP, but using CvD models, for a fixed Kroupa IMF;
 \item[{\it EMILES iT IMF} - ] same as EMILES iT but fixing the IMF slope, \gammab, in Eq.~\ref{eq:method}, to the value obtained from the analysis of optical indices (Paper I), i.e. \gammab$=2.5$ ($1.8$) for the inner (outer)  reference apertures of M31 (see Sec.~\ref{subsec:indices}).
\end{description}
Note that in the first four cases, we assume a Kroupa IMF, in order to compare different methods/models (EMILES and CvD18), while in the last case (EMILES iT IMF), we explore the impact of a { nonstandard} the IMF.

We also add, as extra constraints to Eq.~\ref{eq:method},  the age ($\rm Age_{opt}$) and metallicity ($\rm [Z/H]_{opt}$) estimates obtained from Paper I in the optical spectral range, with their uncertainties, $\rm \sigma_{Age_{opt}}$ and $\rm \sigma_{[Z/H]_{opt}}$, respectively. To this effect, we take the average and standard deviation of age and metallicity estimates  obtained with different { models from Paper I} (see Tab.~\ref{tab:age_met_in_out}), i.e. $\rm Age_{opt}= 8.3 \pm 2$~Gyr ($13.8 \pm 0.4$~Gyr) and $\rm [Z/H]_{opt}=0.3 \pm 0.1$~dex ($-0.1 \pm 0.1$~dex) for the inner (outer) bin.
These values refer to the case of a Kroupa-like IMF. When considering the case of a more bottom-heavy IMF in the center of the bulge, i.e. $\rm \Gamma_b =2.5$ (method ``EMILES iT IMF'', see above), we tend to obtain younger ages (see LB13), and (as a consequence), higher metallicities. Recomputing the age and metallicity estimates as in Tab.~\ref{tab:age_met_in_out}, for $\rm \Gamma_b =2.5$, and for the inner bin, we obtain the constraints $\rm Age_{opt}= 6.0 \pm 2$~Gyr and $\rm [Z/H]_{opt}=0.4 \pm 0.2$~dex, respectively. { Note that the rationale behind the use of extra constraints in Eq.~\ref{eq:method} is that for unresolved stellar populations, the estimate of age and metallicity in the optical has been widely tested in the literature, while the situation in the NIR is less clear, motivating our approach to use the optical results as a benchmark in our analysis. }

\begin{table*}
        \centering
        \caption{Age and metallicity estimates in the center and in the outer region ($R\sim 60$'') of the bulge.}
        \label{tab:age_met_in_out}
        \begin{tabular}{lccccc} 
      \hline
                models & isochrones & $\rm age_{in}$ & $\rm [Z/H]_{in}$ & $\rm age_{out}$ & $\rm [Z/H]_{out}$ \\
\hline
  $\alpha$-MILES & Teramo & $6.6 \pm 1.5$ & $0.4 \pm 0.1$ & $13.2 \pm 0.2$ & $-0.02 \pm 0.02$\\
  EMILES       & Teramo & $11 \pm 1.2$ & $0.25 \pm 0.12$ & $14.0 \pm 0.0$ & $-0.11 \pm 0.03$\\
  EMILES       & Padova00 & $8.3 \pm 0.8$ & $0.22 \pm 0.1$ & $13.8 \pm 0.5$ & $-0.18 \pm 0.04$\\
\hline
        \end{tabular}
\end{table*}

\section{Results}
\label{sec:results}

\subsection{Radial profiles of CO indices}
\label{sec:COrad}

\begin{figure*}
\begin{center}
 \leavevmode
 \includegraphics[width=16cm]{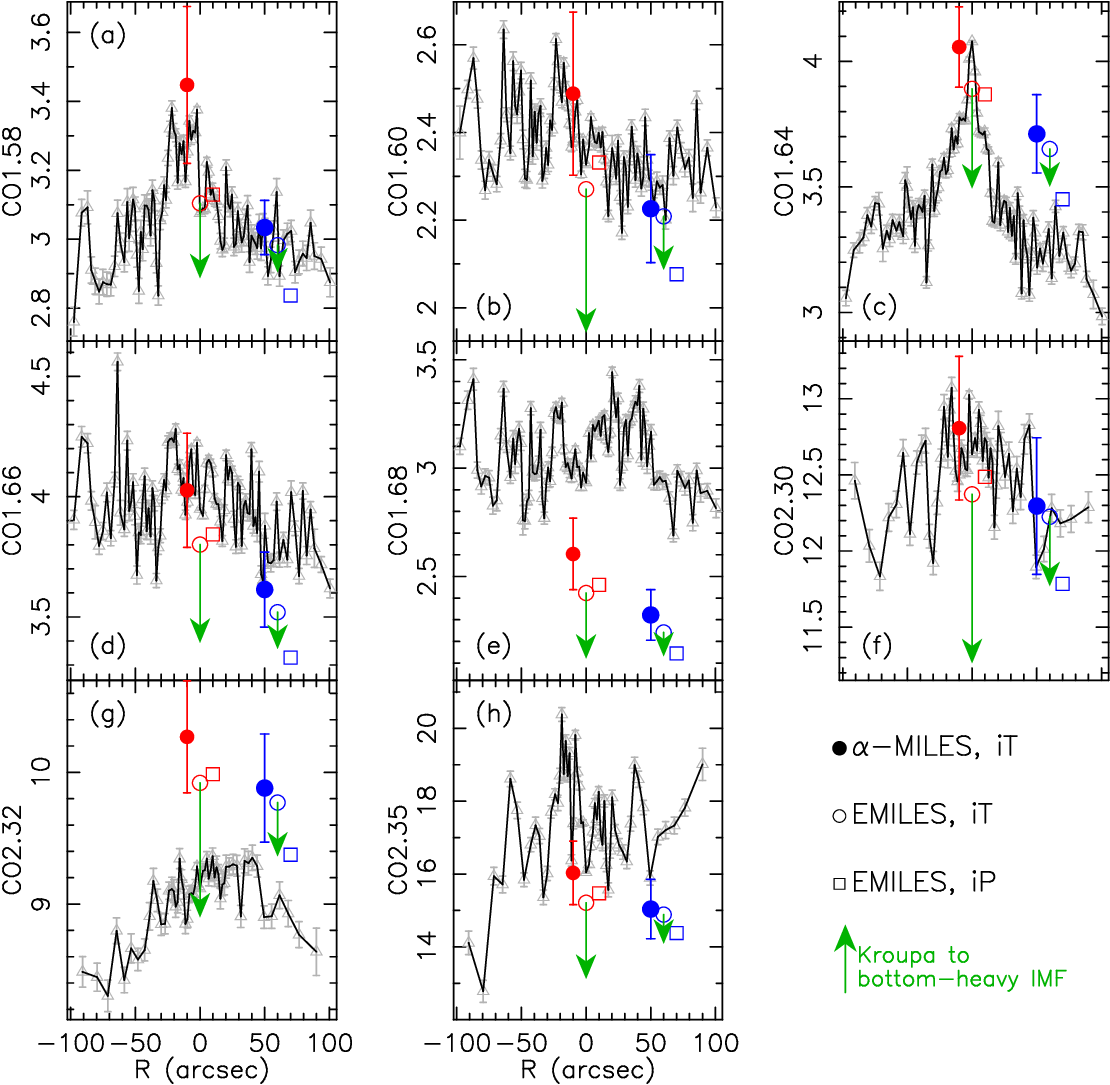}
\end{center}
 \caption{
 Line strengths of H- and K-band CO absorption features are plotted as a function of galactocentric distance, R (in units of arcsec), for the bulge of M31. Negative and positive values of R correspond to opposite sides of the slit. From top to bottom, and left to right, the Figure plots CO absorptions at increasing wavelengths.
 In each panel, measured line strengths are plotted with gray triangles, connected by a black line, and (1--sigma) error bars. The red and blue symbols plot model predictions for an inner ($\rm R \sim$0$''$) and an outer ($\rm R \sim$60$''$) representative aperture, respectively (see the text). Empty circles and squares correspond to Padova and Teramo EMILES 1SSP models (see labels in the lower--right), for a Kroupa-like IMF, with age and metallicity from the optical spectral range (see Paper I). Filled circles are predictions for  Teramo models, with age and metallicity from $\alpha$--MILES models in the optical. Note that, at a given radius, some shift has been applied among models for displaying reasons. The effect of a varying IMF, at fixed age and metallicity, is shown by green arrows. Since CO lines are prominent in giant stars, the arrows point downward in all panels.
 }
   \label{fig:CO_profiles}
\end{figure*}

Figure~\ref{fig:CO_profiles} plots the radial profiles of CO line strengths in the bulge of M31 (see gray symbols, connected by black lines). The observed trends are compared, for illustrative purposes, with predictions of EMILES models based on Padova and Teramo isochrones (empty circles and squares), for the inner and outer reference apertures ($\rm R \sim 0$, and $\rm R \sim 60$'', respectively; see Sec.~\ref{subsec:indices}).
The model predictions are computed for 1SSP models with age and metallicity fixed to the results obtained in Paper I (see Tab.~\ref{tab:age_met_in_out}), from the optical spectral range, for a Kroupa IMF. We also show predictions of EMILES Teramo models, obtained with age and metallicity estimates from $\alpha$--MILES models in the optical spectral range (see filled circles in the Figure). The error bars are reported only for the case of $\alpha$--MILES, and reflect the uncertainties on age, metallicity, and IMF.  The green arrows in the Figure show the effect of changing the IMF only (at fixed age and metallicity), from a Kroupa-like distribution to the solution found in the optical. Note that since CO absorptions are prominent in the spectra of giant stars, a more bottom-heavy IMF implies weaker CO absorptions (see~ELV22a), as shown by the green arrows in the Figure.

As shown in Paper I, most metallic absorption lines  in the optical (e.g., Mgb, Fe, and Na lines), as well as the calcium triplet (CaT) at $\lambda \sim 8600$~\AA,  show steep radial gradients within the inner part ($\sim 30$'') of the M31 bulge, with a flat trend outward. This radial trend follows the radial variation of metallicity inferred from the optical lines (see, e.g., figure~4 of Paper I). In App.~\ref{app:C4668}, we show that the C4668 index, measuring molecular carbon absorption in the optical, exhibits a similar radial behavior as other optical indices. In contrast, Fig.~\ref{fig:CO_profiles} shows that { most NIR CO lines, with the exception of CO1.64, do not show a strong radial gradient in the inner region}. The reason why CO1.64 behaves differently compared to other NIR COs is not clear, although we speculate that this behavior might result from a combined effect of metallicity and other abundance ratios (see Sec.~\ref{sec:fits_abundances}). Note { also that CO2.32 (CO2.30) mildly decreases with radius for galactocentric distances larger than $\sim 50$'' ($\sim 30$'')}, while the radial trends of CO1.58 and CO2.35 show some asymmetries between the two sides of the slit. Although we could not identify any clear sky residual in the regions of CO1.58 and CO2.35 (see also Fig.~\ref{fig:M31spec}), the asymmetries might be due to imperfect sky correction, possibly affecting the two sides of the slit in a different manner, after correcting the spectra for rotation velocity.

Figure~\ref{fig:CO_profiles} also shows that the profiles of all CO features exhibit a scatter that is not accounted for by measurement errors. Since we carefully propagated the error maps through the whole data reduction process (see Sec.~\ref{sec:red}), the scatter in the CO profiles has an intrinsic nature. The origin of this instrinsic scatter will be explored in a forthcoming paper.

Comparing model predictions (circles and squares) and data in Fig.~\ref{fig:CO_profiles},  we see some deviations between predictions from the optical spectral range and  NIR line strengths:
\begin{description}
 \item[{\it Outer region - }] Considering the error bars, the optical solution matches the NIR line strengths at about 1~sigma level, for CO1.58, CO1.60, CO1.66, CO2.30, and CO2.35, respectively, while the agreement is marginal, at $\sim 2$~sigma level, for CO1.64 and CO2.32. The models are inconsistent with the data, being too low (by more than 3~sigma), for CO1.68. We note that these results are independent of the assumed IMF, as in the outer part of the bulge, the effect of a varying IMF is small (see green arrows). The fact that models appear to match the data for most CO lines in the outer region is reassuring, considering that in this region, metallicity is solar, and thus predictions of SSP models are safe.
 \item[{\it Inner region - }] For a Kroupa IMF, model predictions based on the optical spectral range appear to match (at $\sim 1$~sigma) most of the COs, i.e. CO1.58, CO1.60, CO1.64, CO1.66, CO2.30, and CO2.35, while CO2.32 (CO1.68) is above (below) the observed line strengths. In contrast to the outer region, the effect of the IMF is significant in the center, where the bulge of M31 exhibits a (mildly) bottom-heavy IMF  (see Paper I).  As shown by the green arrows in the Figure, the IMF tends to decrease the COs, pushing the model predictions significantly below the observed line strengths. However, one should not swiftly  conclude that this effect implies a tension between CO line strengths and IMF constraints from the optical. Indeed, for a bottom-heavy IMF, one tends to infer young ages, and higher metallicities in the optical. Since COs tend to increase with \zh , the effect of metallicity tends to (partly) balance the effect of the IMF. We further discuss this point in Sec.~\ref{sec:fits}.
 We also note that CO1.68 is systematically offset with respect to the data both in the outer and inner regions. Since we are not able to find any issue with this index related to the data (e.g., sky correction), we conclude that, in general, CO1.68 is poorly modeled, as also confirmed when comparing predictions of different stellar population models (see below).
\end{description}

The fact that some CO lines appear to deviate from models' predictions might be due to (i) uncertainties on stellar population models, especially in the high metallicity regime, and/or (ii) the effect of abundance ratios (that can counteract the IMF effect in the center). We explore points (i) and (ii) in Sec.~\ref{sec:comp_models} and~\ref{sec:fits}, respectively.

\subsection{Predictions of different SP models}
\label{sec:comp_models}

\begin{figure*}
\begin{center}
 \leavevmode
 \includegraphics[width=16cm]{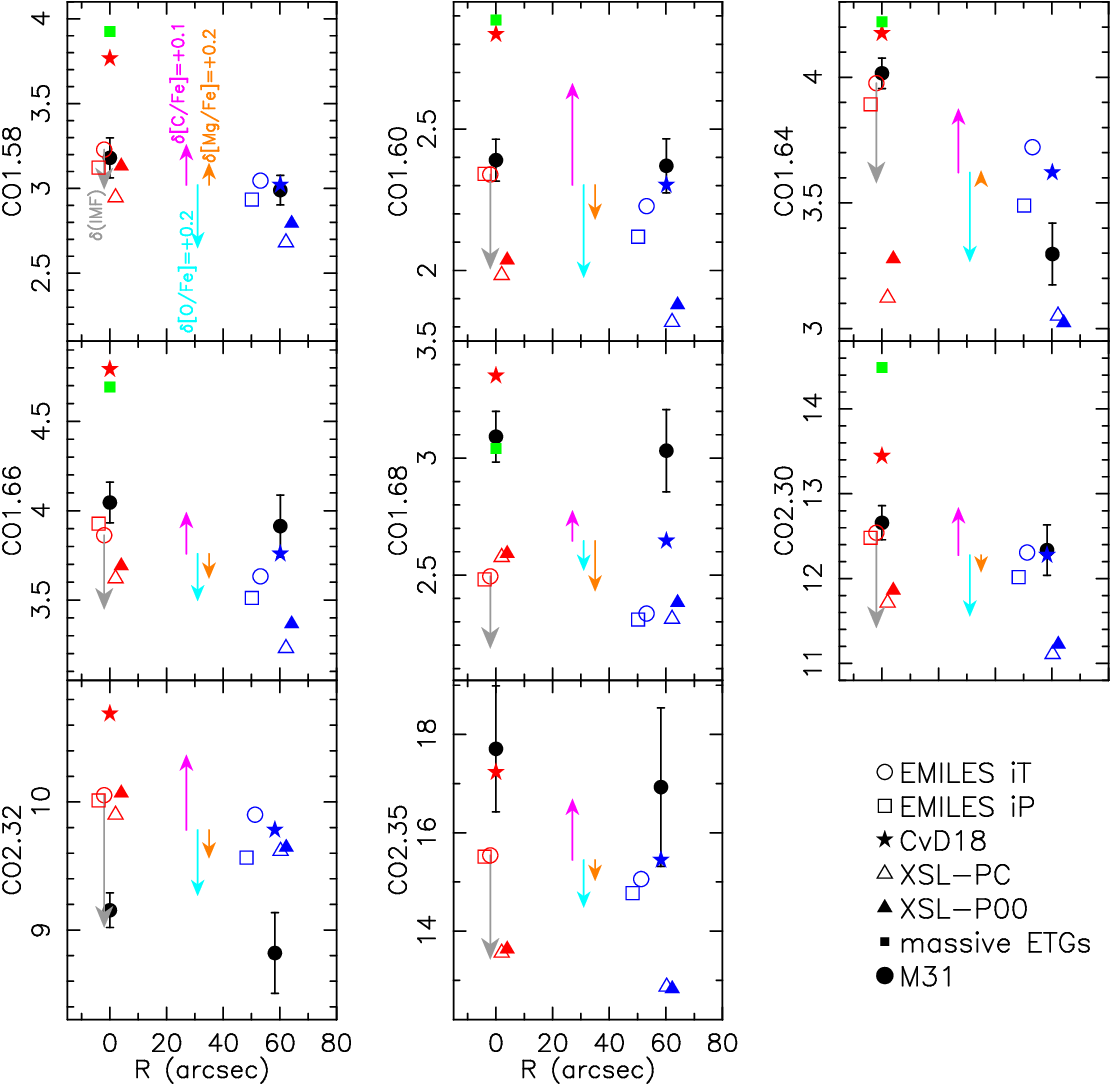}
\end{center}
 \caption{
 Average line strengths of CO absorptions for the inner ($\rm R=0$'') and outer ($\rm R=60$'') reference bins of the M31 bulge (black circles with error bars) are compared to predictions of different stellar population models (see labels in the lower-right corner), namely E-MILES models with Teramo (``EMILES iT'') and Padova00 isochcrones (``EMILES iP''), CvD18 models, as well as XSL models based on Parsec and Padova00 isochrones (XSL-PC and XSL-P00), respectively. Model predictions for the inner (outer) bin, plotted with red (blue) colors, have been computed with 1SSP models, assuming representative values of age and metallicity from the analysis of optical indices (see the text). To perform a fair comparison among models, a Kroupa-like IMF is assumed. The green arrow shows the effect of adopting a more bottom-heavy IMF, as inferred from the optical (see Paper I), for EMILES Padova models. The magenta, cyan, and orange arrows show the effect of increasing \cfe , \ofe , and \mgfe , abundance ratios, at fixed total metallicity (for CvD18 models), by +0.1, +0.2, and +0.2, respectively (see labels in the top--left panel). Note that the average line strengths of massive ETGs,  available for CO2.30 and the other bluer CO indices (see~ELV22a), are systematically higher than most models' predictions and higher than the observed line strengths of the bulge, despite having a more bottom-heavy IMF (see the text).
 }
   \label{fig:CO_comp_models}
\end{figure*}

Figure~\ref{fig:CO_comp_models} compares predictions of different stellar population models to  CO line strengths for the ``inner'' and ``outer'' reference bins of the bulge (see black dots with error bars), obtained by combining spectra from both sides of the slit, as described in Sec.~\ref{subsec:indices} (see also Tab.~\ref{tab:CO_in_out}).
Following the results of Paper I, for the inner (outer) bin, we compute model predictions using 1SSP models, with an age of 9~Gyr ($13.5$~Gyr) and a metallicity of \mh $=0.3$~dex (\mh $=0$~dex). The Figure plots different models with different symbols, as shown from the lower--right labels, namely, EMILES SSPs with Teramo and Padova isochrones, CvD18 models, as well as XSL models with Parsec and Padova isochrones. For a meaningful comparison, all models refer to the same (Kroupa) IMF. At each radial position, predictions from different models are slightly offset, for displaying reasons. Green squares, plotted at $R=0$'', show the average CO line strengths for the sample of seven massive ETGs from~ELV22a (see Sec.~\ref{subsec:XSGs}).
To illustrate the role of abundance ratios, we also show the effect of varying Mg, C, and O, abundances  (see orange, pink, and cyan arrows, at the center of each panel). The index variations for abundance ratios are computed with CvD18 SSP models, at solar metallicity and old age (13.5~Gyr), assuming an enhancement of +0.2~dex for alpha elements (Mg and O; consistent with results of Paper I and~\citealt{Saglia:2010}), and +0.1~dex for C. This choice is based on the results of Sec.~\ref{sec:fits}, where we present fits of optical line strengths and NIR COs, to constrain quantitatively the Mg, O, and C, abundance ratios. Note that the effect of \mgfe , \ofe, \cfe\ in  Figure~\ref{fig:CO_comp_models} is computed at fixed total metallicity~\footnote{To this effect, we compute the variation of total metallicity due to an enhancement of \xfe , and then we decrease \zh\ accordingly, to keep total metallicity fixed. The computation assumes \citet{Asplund:2009} solar abundances. The effect is more important for O, as it gives a substantial contribution to the total metallicity. The computation is performed, separately, for CvD18 and EMILES models. Since EMILES models predict a weaker variation of CO indices with metallicity, the effect of increasing \xfe\ (i.e. the size of the arrows in Fig.~\ref{fig:CO_comp_models}), at fixed total metallicity, is weaker for EMILES, compared to CvD18, models.}, for CvD18 models.
Figure~\ref{fig:CO_comp_models} also shows, with gray arrows, the effect of assuming a bottom-heavy IMF in the center of the bulge (as found in Paper I), rather than a Kroupa-like distribution.

{ Figure~\ref{fig:CO_comp_models} shows that line strengths for the inner and outer apertures are consistent within the error bars for all CO indices, but CO1.64, where a stronger absorption is measured in the inner bin. Note that results are fully consistent with what suggested by direct comparison of the spectra, as seen in Fig.~\ref{fig:spec_in_out}.
Figure~\ref{fig:CO_comp_models} also } highlights several differences among models.
\begin{description}
 \item[{\it Outer region - }] There is good agreement between EMILES and CvD18 models, but for the CO1.68 index, where CvD18 predict higher values than EMILES, approaching, but not reaching, the M31 data (see also Sec.~\ref{sec:COrad}).
 In general, both EMILES and CvD18 models agree with the observed line strengths, while XSL models provide systematically lower COs, with a larger disagreement with respect to the data, for several features (CO1.60, CO1.66, CO2.30, CO2.35). The origin of these offsets is not clear, and cannot be accounted for by varying the abundance ratios (see below).
 Note that the effect of changing the isochrones has a minor impact on CO indices, as seen by the small differences between EMILES iT and iP models, as well as between XSL PC and P00 models.
 \item[{\it Inner region - }] As already shown in Sec.~\ref{sec:COrad}, when assuming a Kroupa-like IMF, EMILES models agree  with observed line strengths for CO1.58, CO1.60, CO1.64, CO1.66 (at $\lesssim 1$ sigma), CO2.30, and CO2.35 (at $\lesssim 2$ sigma), while they tend to underestimate the line strengths of these indices when considering the (mildly) bottom-heavy IMF found in the optical (see the tip of the gray arrows in the Figure). On the contrary, CO2.32 is only matched for a bottom-heavy IMF, while CO1.68 cannot be matched in either case. { In other words, for (solar-scale) EMILES models, there is no solution, accounting only for the effect of age, metallicity, and IMF, as constrained from the optical spectral range, that is able to match all the COs simultaneously.} The XSL models provide lower line strengths, for several CO lines, with respect to both EMILES and CvD18 models, as also found for the outer bulge region (see above), implying an even larger disagreement with the data, compared to EMILES. In contrast to EMILES and XSL, CvD18 models predict a stronger increase of COs  with metallicity, implying higher line strengths, at fixed age and metallicity, in the center (see red stars at $\rm R \sim 0$ in Figure~\ref{fig:CO_comp_models}). It is not clear why CvD18 and EMILES models predict a different dependence on metallicity~\footnote{ At solar metallicity, the updated EMILES models (Sec.~\ref{sec:models}) are very similar to the original EMILES models, with all CO line strengths differing by a few percent. On the other hand, the updated models predict a stronger increase of the COs with metallicity, although significantly weaker than in CvD18 models.}, given that both models adopt similar stellar libraries (IRTF and/or { extended IRTF}). The difference might be due to the  interpolation scheme used to attach stellar spectra to the isochrones, with EMILES adopting a local interpolation, while CvD18 a global approach~\citep{Villaume:2017}.
 The increase of CO line strengths with metallicity suggests that the effect of metallicity and that of a bottom-heavy IMF might be canceling each other in the galaxy center, as they go into opposite directions. However, this scenario does not explain, by itself, the data of massive ETGs (green squares in the Figure). Massive galaxies seem to have a metallicity similar to that of the bulge inner aperture, and a more bottom-heavy IMF (see Sec.~\ref{subsec:XSGs}), but nevertheless, they exhibit significantly higher COs with respect to the bulge. This implies that some other parameter is causing the high CO line strengths observed in massive ETGs (see~ELV22a).
\end{description}


\begin{table}
        \centering
        \caption{Average CO line strengths for the ``inner'' ($R= 0$'') and ``outer'' ($R= 60$'') radial bins of the M31 bulge.}
        \label{tab:CO_in_out}
        \begin{tabular}{lcc} 
          \hline
          index & inner & outer \\
            &  (\AA ) & (\AA ) \\
          \hline 
CO1.58 & $  3.18 \pm  0.12$ & $  2.99 \pm  0.09$ \\ 
CO1.60 & $  2.39 \pm  0.07$ & $  2.37 \pm  0.10$ \\ 
CO1.64 & $  4.02 \pm  0.06$ & $  3.30 \pm  0.12$ \\ 
CO1.66 & $  4.05 \pm  0.11$ & $  3.91 \pm  0.17$ \\ 
CO1.68 & $  3.09 \pm  0.11$ & $  3.03 \pm  0.18$ \\ 
CO2.30 & $ 12.66 \pm  0.20$ & $ 12.34 \pm  0.30$ \\ 
CO2.32 & $  9.15 \pm  0.14$ & $  8.82 \pm  0.32$ \\ 
CO2.35 & $ 17.71 \pm  1.28$ & $ 16.93 \pm  1.61$ \\ 
          \hline
  \end{tabular}
\end{table}

Nonsolar abundance ratios could explain some of the discrepancies between model and observed line strengths seen in Fig.~\ref{fig:CO_comp_models}. Increasing the abundance of alpha elements (both O, and to lower extent, Mg) tends to decrease the COs (with the only exception of CO1.58, that slightly increases with Mg), while carbon abundance shows the opposite trend.  In other terms, at first approximation, the effects of Mg, O, and C tend to cancel out, likely explaining why solar-scale models (EMILES and CvD18) are close to the data of M31 in the outer region, where abundances are not solar.
As discussed by~\citet{CvD12a}, increasing the abundance of alpha elements changes the stellar atmospheric structures in such a way to decrease the COs. On the other hand, enhancing C increases the COs, because the CO molecule has the highest dissociation energy of any known molecule, and the abundance of CO is driven by that of carbon, which is less abundant than O in the stellar atmospheres of old stellar populations.
The effect of other abundance ratios (besides Mg, O, and C) is small for most CO features, as seen by the terms $\rm \sigma_{mod}$ in Tab.~\ref{tab:sigma_mod} (see Sec.~\ref{subsec:fit_indices}), that are smaller than the size of the error bars in Fig.~\ref{fig:CO_comp_models}, and the size of the arrows for varying C and O.  We come back  discussing this point in Sec.~\ref{sec:fits_abundances}.
We note that, given the low COs predicted by XSL models, in order to match the M31 data with these models, one should invoke a [C/Fe] abundance up to $\sim 0.3$~dex in the outer region (see, e.g., the size of the arrows for CO2.30), which is in disagreement with the optical constraints, as shown in the following section.

\subsection{Fitting optical and CO indices}
\label{sec:fits}
Since CO indices are affected by several stellar population parameters, such as metallicity, abundance ratios, and IMF, we try to match the observed line strengths quantitatively, with different fitting methods, as described in Sec.~\ref{subsec:fit_indices}.
 Tab.~\ref{tab:fits_ews} reports the best-fitting results, i.e. the abundance ratios  of \cfe, \mgfe, and \ofe, as well as the reduced chi-square, $\rm \chi^2_{\nu}$, of the fits, for the inner and outer reference apertures.
Discrepancies between input quantities (i.e. age and metallicity estimates from Paper I, as well as observed line strengths) and best-fitting parameters, in units of sigma, are shown in Fig.~\ref{fig:disc_ews}, while Fig.~\ref{fig:cmgo} plots \cfe, \ofe, and \mgfe, for different fitting cases (x-axis), for the inner and outer apertures (see top and bottom panels, respectively).

\subsubsection{Optical indices}
\label{sec:fits_optical}
All models match the optical indices very well , as shown by the low values of the reduced $\chi^2$ in Tab.~\ref{tab:age_met_in_out}, with discrepancies within $\sim 1$~sigma for all input quantities (see black lines in Fig.~\ref{fig:disc_ews}). This is  in  agreement with what already found by Paper I for a different set of optical spectral features.

In the center of the bulge, the abundance ratio of Mg is $\sim 0.14$~dex, while that of O tends to be lower ($0.02 \pm 0.08$), but still consistent with \mgfe\ within the error bar. Note that the larger error bar on \ofe\ is due to the fact that O abundance is difficult to constrain using optical indices only~\citep{JTM:2012}. Carbon abundance is lower than that of Mg, with \cfe$=0.04 \pm 0.02$. We verified that assuming a bottom-heavy IMF in the center (i.e. $\rm \Gamma_b=2.5$, see Paper I) does not change significantly these conclusions.

In the outer region, we find \mgfe$\sim 0.16$~dex, and a larger value of \ofe\ ($\sim 0.27$~dex).
These values are consistent with the
estimates of $\rm [\alpha/Fe]$ ($\sim 0.2$~dex) from Paper I and S10, although one should note that here we are fitting  the abundances of Mg and O individually, rather than assuming all $\alpha$ elements to track each other.
The abundance of carbon ($0.14 \pm 0.02$) tends to be consistent with that of Mg, and significantly larger compared to that for the inner bin (\cfe$=0.04 \pm 0.02$).

We verified that the above results are robust against the set of indices fitted in the optical. For instance, considering  the outer bin and including also the indices \tioi, \tioiir, \fef, \cnt, \cno, \caf, and \cat\ (see Paper I and references therein for the definition of these indices), as well as \tife, \nfe, and \cafe\ abundance ratios in the fit, gives \cfe=$0.14 \pm 0.02$~dex, \mgfe=$0.19 \pm 0.02$~dex, and \ofe=$0.28 \pm 0.04$~dex, consistent with results reported in Tab.~\ref{tab:age_met_in_out}. This confirms that in the outer bin, O is actually enhanced with respect to Mg and C, as also shown in the following section, from the analysis of CO indices.

\subsubsection{CO and optical indices}
\label{sec:fits_CO}
Including CO line strengths produces a higher { reduced chi-square ($\chi_{\nu}^2$)} compared to the optical fits (Tab.~\ref{tab:age_met_in_out}). 
For $\nu=10$ degrees of freedom (as it is the case for fits including CO indices), the probability to obtain a reduced chi--square larger than the given value, is $5 \%$ ($1 \%$) for $\chi_{\nu}^2 \sim 1.83$ ($2.12$), meaning that we should reject solutions with $\chi_{\nu}^2 \gtrsim 2$. Hence, all results in Tab.~\ref{tab:fits_ews} are statistically acceptable.

In the outer bin, optical and CO indices are well fitted for all methods. In particular, all COs are matched at the $2$--$3$~sigma level, with the largest deviations for CO1.64 and CO2.32. We further discuss these deviations in Sec.~\ref{sec:fits_abundances}. We stress out that matching CO lines mostly results from a balance between C and O abundance ratios. This requires, for all models, a stronger enhancement of O with respect to C, as shown in Fig.~\ref{fig:cmgo}. Note also that the values of \cfe\ and \mgfe\ ($\sim 0.15$~dex) are fully consistent with those found in the optical. 

Fig.~\ref{fig:disc_ews} shows that even for the inner bin we are able to match all COs at the $2$~sigma level. All models tend to overestimate, by $\sim 2$~sigma, the CO2.32 index, as also found for the outer bin, and to lower extent also the CO1.58 index. Indeed, we verified that excluding CO2.32 from the fits does not change significantly our results~\footnote{The central passband of CO2.32 is affected by a strong telluric line at $\lambda \sim 23164$~\AA . However, this appears to be well corrected in our data, with no significant residual { but in a narrow region between the blue pseudocontinuum and the central passband of the index (see Fig.~\ref{fig:spec_in_out})}.}. The estimates of \mgfe, \cfe, and \ofe\ agree with results from the optical, within the error bars, although the fit with CvD18 models tends to provide lower \mgfe\ ($\sim 0.08$~dex) compared to EMILES (see Tab.~\ref{tab:fits_ews}), and therefore, more consistent \mgfe\ and \ofe\ abundance ratios. Note that although EMILES and CvD18 models provide a very different behavior of CO indices with metallicity (see Sec.~\ref{sec:comp_models}), one can obtain a good fit with both sets of models. The reason is that in order to fit the (optical and NIR) data, CvD18 models require significantly lower metallicity (\zh $\sim 0.12$~dex), and older ages, compared to EMILES models (\zh $\sim 0.3$~dex).

To summarize, we find that we can match reasonably well the CO features both in the center and the outer aperture of the bulge. C and Mg track each other in the outer bulge, with an overabundance of $\sim 0.15$~dex, while O tends to be above Mg (and C). In the center, the abundance of C is about half (or even lower) than that in the outer region. Mg appears to be higher than C, and more consistent with results for the outer bin, while O is consistent with both Mg and C, given the error bars.

\subsubsection{Na and other abundance ratios}
\label{sec:fits_abundances}
CO indices are mostly sensitive to C and O abundance ratios (Sec.~\ref{subsec:fit_indices}). Therefore, we do not expect significant changes to our results when including the effect of other abundance ratios in the analysis. We verified this point by repeating the fit for the inner bin, using EMILES Teramo models, and adding in turn, the abundance of \sife, \tife, \cafe,  \nfe, and \nafe\ as an extra free parameter in our fitting procedure~\footnote{In each case, the values of $\rm \sigma_{mod}$ (see Tab.~\ref{tab:sigma_mod}) were updated according to the abundance ratios included in the fit. For instance, when including \tife\, we computed
$\rm \sigma_{mod}$ as the rms of differences between line strengths for CvD18 models with enhanced Ca, Si, Na, N, and K (i.e. excluding Ti), and that of the CvD18 solar-scale model. Note that when fitting \nafe , we set $\rm \sigma_{mod}=0.08$~\AA\ for CO2.32, rather than assuming a larger value of $0.5$~\AA\ (see Sec.~\ref{subsec:fit_indices}). }. We found best-fitting values of \cfe\ in the range from $0.03$ to $0.05$~dex, \mgfe\ in the range from $0.12$ to $0.16$~dex, and \ofe\ in the range from $0.04$ to $0.12$~dex, consistent with results reported in Tab.~\ref{tab:fits_ews}.
A relevant effect on (some) CO features is that of Na abundance. Na is one of the main electron donors in the stellar atmospheres, with its abundance affecting also the strengths of other absorption features \citep{CvD12a}. Moreover, \nafe\ is significantly high in the bulge, particularly in its center (see Paper I).
According to both CvD18 and Na-enhanced EMILES models, the K-band COs -- in particular CO2.32 and CO2.35~\footnote{We note that for both CO2.32 and CO2.35, there is a contribution of Na absorption in the pseucontinuum of the features, as shown in figure~A3 of EVL21.} -- are significantly affected by \nafe , in contrast to  the H-band COs and CO2.30. Moreover, all COs are predicted to (slightly) decrease with increasing \nafe , with the exception of CO1.64, that tends to increase with \nafe . This could (partly) explain why the index has a peaked profile in the galaxy center, in contrast to the other COs (see Sec.~\ref{sec:COrad}). More in detail, using CvD18 models (with age of 13.5Gyr and solar metallicity), for $\rm \delta[Na/Fe]=+0.6$~dex, we find $\rm \delta CO2.32 \sim -0.5$~\AA , $\rm \delta CO2.35 \sim -1$~\AA , and $\rm \delta CO1.64 \sim +0.2$~\AA.
Interestingly, when fitting for \nafe , without including the Na lines, we find a best-fitting Na abundance of $0.6 \pm 0.17$, consistent with results obtained in the optical (see figure~7 of Paper I).

\subsubsection{Fits with different IMFs}
\label{sec:fits_IMF}
\begin{figure}
\begin{center}
 \leavevmode
 \includegraphics[width=8cm]{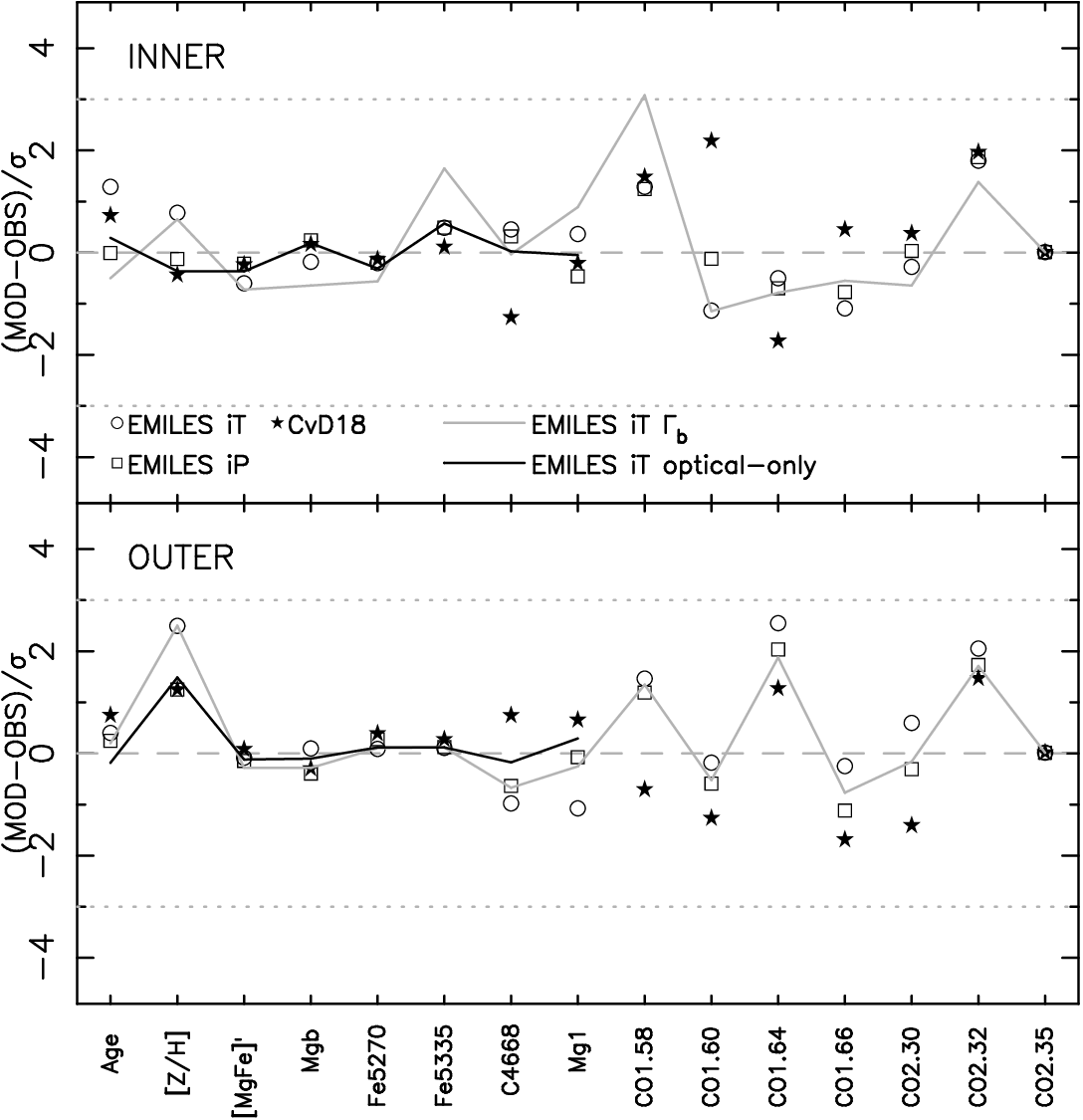}
\end{center}
 \caption{
 Discrepancy of best-fitting and observed line strengths, in units of $\rm \sigma$ (see Eq.~\ref{eq:method}), for different fitting methods (see the labels in the lower part of the upper panel).
 }
   \label{fig:disc_ews}
\end{figure}

The gray lines in Fig.~\ref{fig:disc_ews} show the discrepancy between model and observed indices when assuming the IMF slope derived from the optical spectral range, i.e. \gammab$=2.5$ ($1.8$) for the inner (outer) radial bin (see Paper I). Here, we focus on results  based on EMILES Teramo models.
In the outer bin, the quality of the fit is slightly better for \gammab$=1.8$, with a lower reduced chi--square ($\rm \chi^2_{\nu} \sim 1.34$) with respect to the case of a Kroupa-like distribution ($\rm \chi^2_{\nu} \sim 1.87$; see Tab.~\ref{tab:fits_ews}). On the contrary, for the inner bin, the fit is worst for a bottom-heavy IMF (\gammab$=2.5$), with $\rm \chi^2_{\nu} \sim 1.94$, compared to the case of a Kroupa-like IMF ($\rm \chi^2_{\nu} \sim 1.07$). We note that a value of $1.94$ is still acceptable, considering the number of degrees of freedom (see Sec.~\ref{sec:fits_CO}). Moreover, a closer inspection of Fig.~\ref{fig:disc_ews} reveals that the chi-square is driven mostly by one index (CO1.58), showing a larger deviation (see gray line in the top panel), compared to the case of a Kroupa-like IMF (empty circles).
Including the effect of other abundance ratios does not change significantly our results (Sec.~\ref{sec:fits_abundances}). As already discussed in LB13, for a bottom-heavy IMF, one tends to infer younger ages, and therefore higher metallicity, compared to a Kroupa-like distribution. For \gammab$=2.5$, and EMILES Teramo models, one infers (1SSP) ages as young as 5~Gyr in the central bin of M31, with metallicities as high as $\sim $0.5~dex~\footnote{ Note that this requires a significant extrapolation of the models in the super-solar metallicity regime, compared to the case of a Kroupa-like IMF, where the inferred metallicity is $\sim$0.3~dex.}. The high metallicity compensates the effect of the IMF that pushes the CO line strengths to lower values (see Fig.~\ref{fig:CO_comp_models}). This explains why we are able to match the COs both for a Kroupa and a (mildly) bottom-heavy distribution.

\begin{figure*}
\begin{center}
 \leavevmode
 \includegraphics[width=10cm]{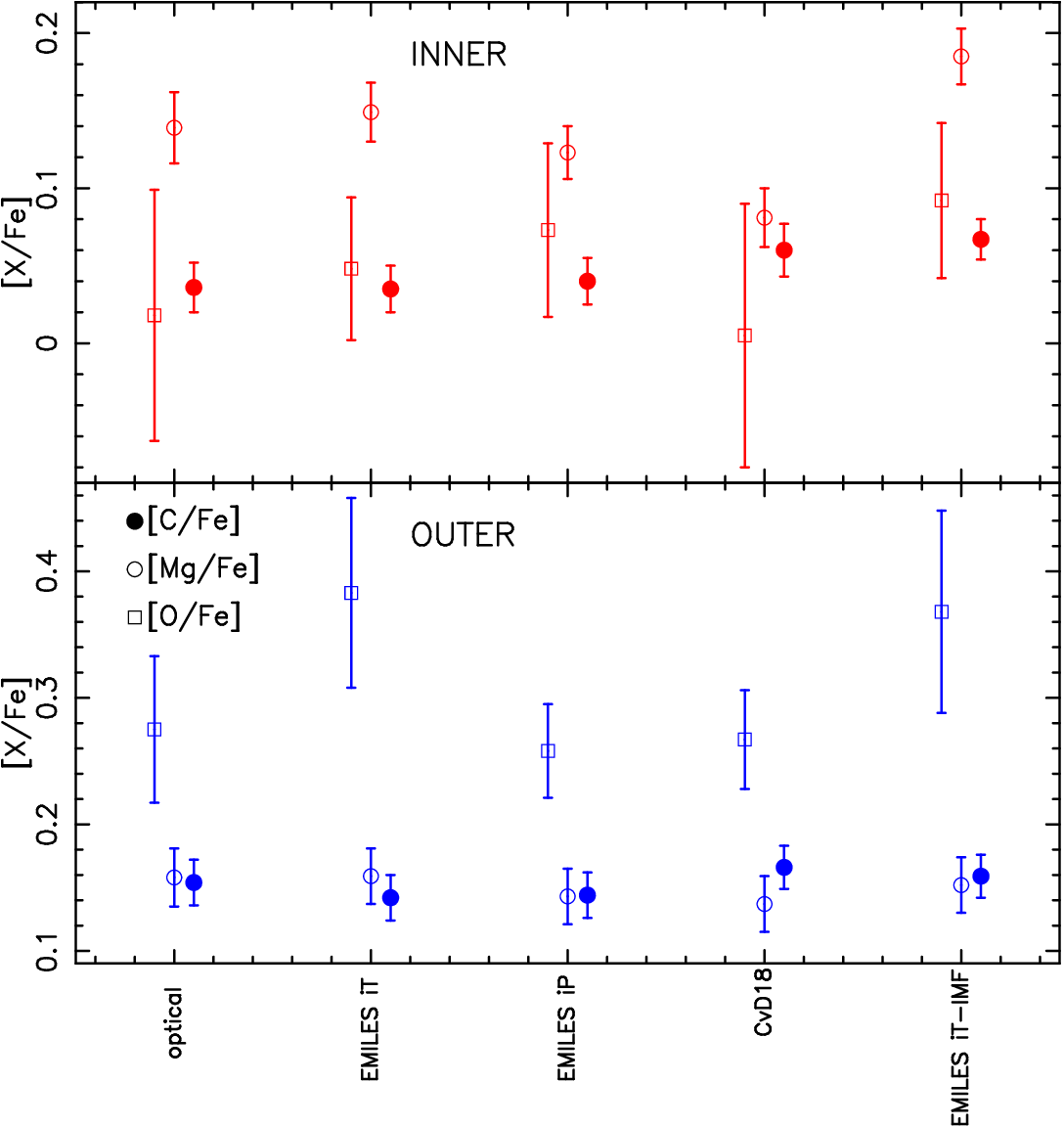}
\end{center}
 \caption{
 \cfe, \mgfe , and \ofe\ abundance ratios (dots, empty circles, and squares, respectively) are shown for different methods and sets of indices (see the labels on the x-axis), and for the inner (top) and outer (bottom) radial bins of the M31 bulge.
 Error bars correspond to 1-sigma uncertainties.
 }
   \label{fig:cmgo}
\end{figure*}

\subsection{Uncertainties on optical-NIR best-fitting parameters}
\label{sec:deg}

\begin{figure*}
\begin{center}
 \leavevmode
 \includegraphics[width=16cm]{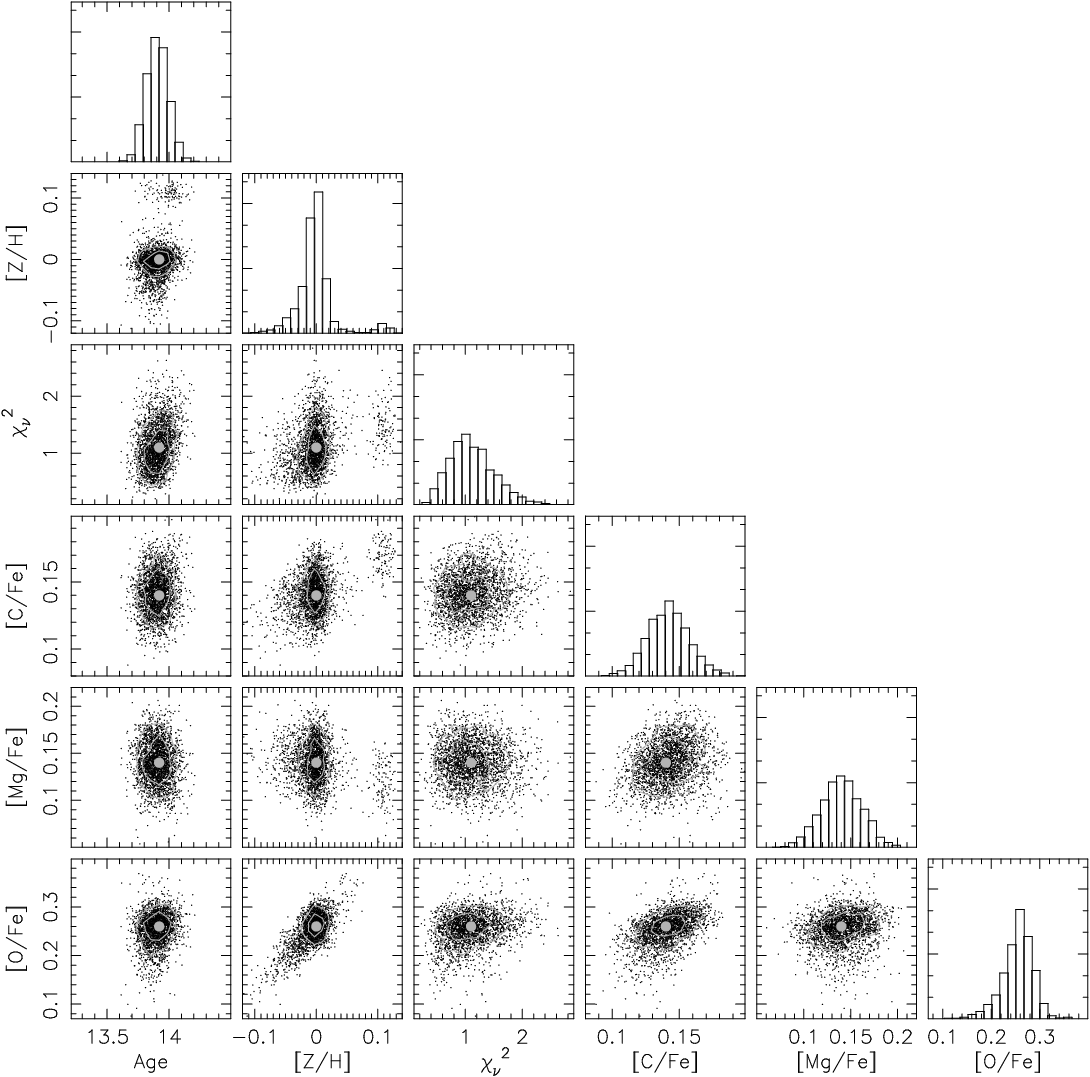}
\end{center}
 \caption{ Corner plot showing the distribution of the best-fitting parameters, namely the age, metallicity, \cfe, \mgfe, \ofe, as well as reduced chi-square, $\chi_\nu^2$, for the outer aperture of the M31 bulge, when fitting both optical indices and CO lines, with EMILES Padova models (method EMILES iP, see Sec.~\ref{subsec:fit_indices}). For each panel that plots a pair of different parameters, black points correspond to results for different bootstrap iterations, where observed line strengths are randomly shifted according to their uncertainties; gray curves are iso-density contours, with density levels corresponding to 1- and 2-sigma confidence contours for a two-dimensional random deviate; big gray dots are the best-fitting solutions, for the nominal values of line strengths.
 }
   \label{fig:corner}
\end{figure*}

{
In order to discuss possible degeneracies among best-fitting stellar population parameters in our analysis, we show in Fig.~\ref{fig:corner} the distribution of best-fitting solutions for the outer aperture of the M31 bulge, in the case where both optical and CO indices are fitted simultaneously, with EMILES Padova models (method EMILES iP, see Sec.~\ref{subsec:fit_indices}). Different points in the Figure correspond to results for different bootstrap iterations, where observed line strengths are shifted according to their uncertainties, assuming Gaussian distributions. The Figure shows that all parameters are well constrained, with some correlated variation only seen, for some solutions,  in the \ofe--\zh\ plane. This is somehow expected, as at fixed total metallicity, both CO lines as well as other optical indices, such as \mgfep\ and Mg1, tend to decrease with \ofe , while increasing with \zh. However, \zh\ is well constrained in the optical spectral range (see Paper I), with values  consistent with previous works~\citep{Saglia:2010, Saglia:2018} as well as with resolved stellar population studies~\citep{SJ:2005}. Indeed, assuming no priors on age and metallicity (i.e. removing the age and metallicity contraints in Eq.~\ref{eq:method}) would lead to best-fitting results very similar to those reported in Fig.~\ref{fig:corner}, but with an enhanced tail of correlated solutions in the \ofe--\zh\ plane. Similar findings apply to the inner aperture of the M31 bulge.
We note that the correlated variation of \zh\ and \ofe\ does not affect significantly our conclusions, since our main aim in the present work is to analyze if optical predictions are consistent with CO lines. }

\section{Discussion}
\label{sec:discussion}
\subsection{Abundance ratios}
We have derived \mgfe, \ofe, and \cfe\ abundance ratios for the center ($\rm R < 3$'') and the outer region ($\rm 35'' \! < \! R \! < \! 100$'') of the bulge of M31. Both regions are part of the classical bulge of Andromeda~\citep{BD:2017}. Since the profiles of optical and NIR line strengths are flat outside a radius of $\sim 30$'' (i.e. $\sim 100$~pc; see Paper I), our outer region describes the bulk properties of the classical bulge. In agreement with S10 and Paper I, we find that the majority of the bulge is enhanced in Mg, with \mgfe$\sim 0.15$~dex. We also find that O is overabundant with respect to Mg,  by $\sim 0.1$~dex (see Tab.~\ref{tab:fits_ews}), while the abundance of C is consistent with that of Mg (\cfe $\sim 0.15$~dex). On the other hand, within a few arcsec of the center, Mg remains constant, or slightly decreases (\mgfe $\sim 0.08$--$0.19$~dex, depending on the method), with respect to the outer region, \cfe\ is about half of the outer value ($\sim 0.05$~dex), while O is found to follow C, but with a large error bar.
However, one should note that all our abundance estimates are relative to the abundance pattern of stars used to construct the SSP models. While at solar metallicity, stars in the solar neighborhood have both \mgfe\ and \ofe\ around zero~\citep{Bensby:2014}, \cfe\ is slightly negative, by a few percent~\citep{Bedell:2018, Franchini:2020}). Since this pattern is not corrected for in the models, our \cfe\ abundances might be slightly overestimated, with true \cfe\ values being around zero ($\sim 0.1$~dex) for the bulge center (outer bin). This is important in light of the comparison to massive ETGs (see below). Moreover, in the super-solar metallicity regime, \ofe\ is  significantly negative for Milky-Way stars, being around $-0.2$~dex at $\rm [Fe/H] \sim 0.3$~dex (see figure~15 of~\citealt{Bensby:2014}). Therefore, in the center of the bulge, the true \ofe\ should be similar, or even higher, than \mgfe . In other words, after correcting for the MW abundance pattern, both the center and the outer region of the bulge appear to be alpha--enhanced, in both Mg and O, while C abundance decreases significantly in the center.

Our results can be compared with the chemical evolution model of~\citet[hereafter MU14]{MU:2014}, who found that, in order to reproduce the metallicity distribution of individual stars in the bulge of M31~\citep{SJ:2005}, this should have formed very quickly, with a flatter IMF than in the solar neighborhood.
In the MU14 models, stars with a total metallicity of $\rm [M/H] \sim 0$ (i.e. $\rm [Fe/H]$ about $-0.15$~dex, see their figure~2) have
\mgfe$\sim 0.35$~dex, a slightly higher oxygen abundance  (\ofe$\sim 0.4$~dex), and \cfe$ \sim 0.2$~dex. While the prediction  \ofe $>$\mgfe\ is qualitatively consistent with our results (see bottom panel of Fig.~\ref{fig:cmgo}), we find lower values of \mgfe\ and \cfe\ ($\sim 0.15$~dex), possibly implying that the bulge did not experience such a quick formation as in the MU14 model, and/or did not formed with a top-heavy IMF. Detailed chemical evolution models would be required to address this issue.

Our abundance determinations can be also compared with estimates for the Galactic bulge. At solar metallicity, stars in the MW bulge have \mgfe\ around zero, or slightly super-solar~\citep{Bensby:2017}, with \ofe\ and \cfe\ being consistent with zero ( \citealt{Bensby:2021}; see also~\citealt{Barbuy:2018, Queiroz:2021}). Hence, the bulge of M31 is significantly more enhanced in Mg, O, and C, with respect to the bulge of the Milky-Way (MW). This might be due to the fact that the classical bulge of M31 has formed over a shorter timescale with respect to the MW bulge, with the latter mostly consisting of disk-like material, formed through secular evolution~\citep{WG:2013}. This point illustrates the importance of  performing a detailed determination of chemical abundance ratios in the bulge of M31, to shed light on the formation history of (classical) bulges.

\subsection{COs in the M31 bulge and massive galaxies}
As discussed in~ELV22a and also shown in the present paper, massive ellipticals have CO line strengths significantly stronger than predictions of state-of-the-art stellar population models. On the contrary, we find that the bulge of M31, despite its old age and high central metallicity, similar to massive ellipticals, exhibits significantly lower COs, close to models' predictions.
We discuss here possible explanations of the difference between the bulge and massive galaxies.

\begin{description}
\item[{\it Intermediate-age populations} - ]
Strong COs might be due to the contribution of AGB stars from intermediate-age populations (1~Gyr or younger). At these ages, AGB dominate the infrared flux of galaxies, while at older ages the NIR flux is dominated by the longer-lived RGB stars (see~\citealt{Salaris:2014}, and references therein).
Indeed, studying the CMD diagram from HST observations, \citet{Dong:2018} found that the majority of stars in the M31 bulge are old, with a few percent contribution from intermediate-age populations ($<2$~Gyr), the latter increasing toward the bulge center. This is consistent with the finding of a younger age in the inner region of the bulge (see Paper I). However, we find similar CO line strengths for the inner and outer regions of the bulge, significantly lower than those of massive ellipticals. This reinforces the idea that intermediate-age populations do not drive CO absorptions, in agreement with the conclusions of~ELV22b, who analyzed CO lines for the massive relic galaxy NGC\,1277, finding line strengths as strong as for other massive galaxies.
 \item[{\it Metallicity} - ] The center of the M31 bulge has a metallicity, \zh, between 0.1 and 0.4~dex, depending on the stellar population model adopted to derive it (see Tab.~\ref{tab:age_met_in_out} and Sec.~\ref{sec:fits_CO}). \citet{Saglia:2018} also found a super-solar metallicity, $\sim 0.35$~dex, in the center of the bulge. The metallicity estimate is even higher when accounting for a variable IMF (Sec.~\ref{sec:fits_IMF}). On the other hand, the sample of massive galaxies from~\citet{LB:19} has a central \zh\ between 0.15~dex and 0.4~dex (see their figure~4),  i.e. in a similar range as the M31 bulge. Although we cannot exclude, given the current uncertainties, that massive galaxies have higher metallicity than the center of the M31 bulge (by $\sim 0.1$--$0.2$~dex), this difference is likely too small to account for the difference of CO line strengths in Fig.~\ref{fig:CO_comp_models}.
 Hence, metallicity alone cannot explain the difference between massive galaxies and the M31 bulge. It remains to be understood the true dependence of CO lines on \zh . Both EMILES and XSL models predict a mild dependence on metallicity, while CvD18 SSPs exhibit a strong increase of CO line strengths with \zh . This is surprising, considering that the version of EMILES models used in the present work also includes the same stellar library ({ extended IRTF}) as CvD18 models. Note that the best-fitting metallicity in the center of the bulge is lower for CvD18 than other models. Therefore, in practice, all models are able to fit the observations. Studying CO lines for larger samples of stellar systems, spanning a wide metallicity range, may help to shed light on this issue.
 \item[{\it IMF shape} - ] Since CO features are stronger in giant stars, than in dwarfs, a more bottom-heavy IMF makes the COs weaker. Surprisingly, although massive galaxies have a more bottom-heavy IMF than the bulge of M31, they exhibit stronger, not weaker, CO line strengths (see Fig.~\ref{fig:CO_comp_models}).  In principle, for an old population, one can increase the COs without affecting significantly the fraction of low-mass stars, by changing the slope of the IMF at $\sim 1$~$\rm M_{\odot}$. This mass corresponds to stars still alive in an old stellar population, but already evolved off the main sequence, in the RGB and AGB phases. These late-type giants have strong COs~(see ELV22a). To explore this possibility, we start considering an SSP, with super-solar metallicity ([Z/H]$=+0.26$), an age of 10~Gyr, and a single power-law IMF of slope  $\Gamma=2.3$, i.e. a bottom-heavy distribution~\footnote{In this notation, $\Gamma=1.35$ corresponds to the Salpeter IMF}. This model has NaI8200=$0.9$~\AA, and $\rm CO2.30=10.5$~\AA. Note that the NaI8200 is strongly sensitive to the presence of low-mass stars, and a value of $0.9$~\AA\ matches the observations for massive galaxies, having a bottom-heavy IMF~\citep{LB:13, Spiniello:2014}.
 On the contrary, a value of $\rm CO2.30=10.5$~\AA\ (we focus on this index for illustrative purposes) is far too low (by $\sim 4$~\AA ) compared to that of massive ETGs (see Fig.~\ref{fig:CO_comp_models}). Hence, we recompute the same model with a flat slope ($\Gamma=0$) for masses $\ge 1$~$\rm M_{\odot}$. We find a small change in NaI8200 (about $-0.05$~\AA), and an increase of $\sim 0.5$~\AA\ in CO2.30. Although this variation of CO2.30 is significant (compared to the error bars in Fig.~\ref{fig:CO_comp_models}),  it is far too small to match the observed line strengths for massive galaxies ($\rm CO2.30 \sim 14.5$~\AA ). We conclude that an IMF shape which is bottom-heavy at low mass, {\it and} top-heavy at high mass cannot explain the observations.
 \item[{\it Abundance ratios} - ] Massive ETGs and the center of M31 differ significantly in their abundance ratios. Massive galaxies are significantly more enhanced in $\alpha$ elements (\mgfe$\sim 0.35$, see~\citealt{LB:19}) and carbon (\cfe$\sim 0.2$dex, see~\citealt{CGvD:2014}), compared to the center of the M31 bulge, where we find  \cfe\ slightly above solar (see above), \mgfe$\sim 0.15$~dex, and \ofe\ consistent, or slightly above, \mgfe.
 The difference in the abundance of alpha elements is likely not driving the difference in COs, as O overabundance tends to decrease CO line strengths (at fixed total metallicity), while Mg has a minor effect  on them. On the other hand, C abundance tends to increase the COs, and therefore it is the most natural candidate to explain the difference between massive galaxies and the bulge. However, in order to explain the observations, one should invoke a significantly stronger sensitivity of CO lines to \cfe, in the super-solar metallicity regime, compared to what predicted by CvD18 models. This might actually be the case, as in the NIR, the response to abundance ratios is driven by low-temperature stars, whose modeling is notoriously uncertain~(see ELV22a).
\end{description}

\section{Summary and conclusions}
\label{sec:summary}

Based on new H- and K-band spectroscopy for the bulge of M31, we have studied several CO absorption features (CO1.58, CO1.60, CO1.64, CO1.66, CO1.68, CO2.30, CO2.32, and CO2.35) out to a galactocentric distance of $\sim 100 ''$ ($\sim 400$~pc). The results of the present work can be summarized as follows:
\begin{description}
 \item[ - ] { Most COs, except for CO1.64, do not show  significant radial gradients, in contrast to most metallic lines in the optical spectral range}.
 \item[ - ] The observed COs are close to predictions of EMILES and CvD18 stellar population models, while XSL models tend to underestimate the observed line strengths. The index CO1.68 is significantly under-predicted by all models (with a lower discrepancy for CvD18 SSPs).
 \item[ - ] For both EMILES and XSL models, SSPs based on different isochrones show very similar CO line strengths, implying that the treatment of different stellar evolutionary phases cannot explain differences between models and data.
 \item[ - ] The lack of radial gradients in most COs is due to the competing effect of metallicity and carbon abundance, pushing the indices to higher values, and that of an IMF and oxygen overabundance, which tends to decrease the indices in the center of the bulge. The behavior of CO1.64 is likely explained by the effect of a Na abundance.
 \item[ - ] We were able to match, simultaneously, optical indices and all COs (with the exception of CO 1.68), by including the age, metallicity, C, Mg, and O abundance ratios as free-fitting parameters, and by imposing the constraints from the optical spectral range.
 \item[ - ] The majority of the M31 bulge (our ``outer'' bin) is alpha enhanced, consistent with previous works. By fitting individual abundances, we find \mgfe$\sim 0.15$~dex, \ofe\ slightly above \mgfe\ (by $\sim 0.1$~dex), and a C abundance consistent with that of Mg. In the central (few arcsec) region, we still find an enhancement of O and Mg, with a significantly lower \cfe\ (slightly above, or consistent with, solar composition).
 \item[ - ] The CO line strengths of the M31 bulge are significantly lower than those of massive ETGs. We speculate that this difference might arise because of a difference in the C abundance, as well as, to some extent, the total metallicity.
\end{description}

The emerging picture is that CO line strengths result from a complex interplay between the metallicity, IMF, and abundance ratios (mostly C and O). Our work points to the need of further improving state-of-the-art stellar population models, and theoretical responses to varying abundance ratios, in the NIR spectral range, especially in the super-solar metallicity regime.

\begin{acknowledgements}
{ We thank the anonymous referee for his/her helpful comments, that significantly helped us to improve our manuscirpt.}
F.L.B. and A.P. acknowledge support from the grant PRIN-INAF-2019 1.05.01.85. E.E. and A.V. acknowledge support through the IAC project TRACES which is partially supported through the state budget and the regional budget of the Consejer\'\i a de Econom\'\i a, Industria, Comercio y Conocimiento of the Canary Islands Autonomous Community.
F.L.B. acknowledges support from the INAF grant 1.05.23.04.01.
The present paper is based on LUCI-LBT spectroscopy acquired under LBTB and LBTI time. The LBT is an international collaboration among institutions in the United States, Italy and Germany. LBT Corporation Members are: The University of Arizona on behalf of the Arizona Board of Regents; Istituto Nazionale di Astrofisica, Italy; LBT Beteiligungsgesellschaft, Germany, representing the Max-Planck Society, The Leibniz Institute for Astrophysics Potsdam, and Heidelberg University; The Ohio State University, representing OSU, University of Notre Dame, University of Minnesota and University of Virginia. The data for massive ETGs have been acquired with ESO Telescopes at the Paranal
Observatory under programmes ID 092.B-0378, 094.B-0747, 097.B-0229 (PI: FLB).
\end{acknowledgements}

%
%

\begin{appendix} 

  \section{Kinematics of the bulge in the H and K bands}
\label{app:kin}
\begin{figure}
\begin{center}
 \leavevmode
 \includegraphics[width=8cm]{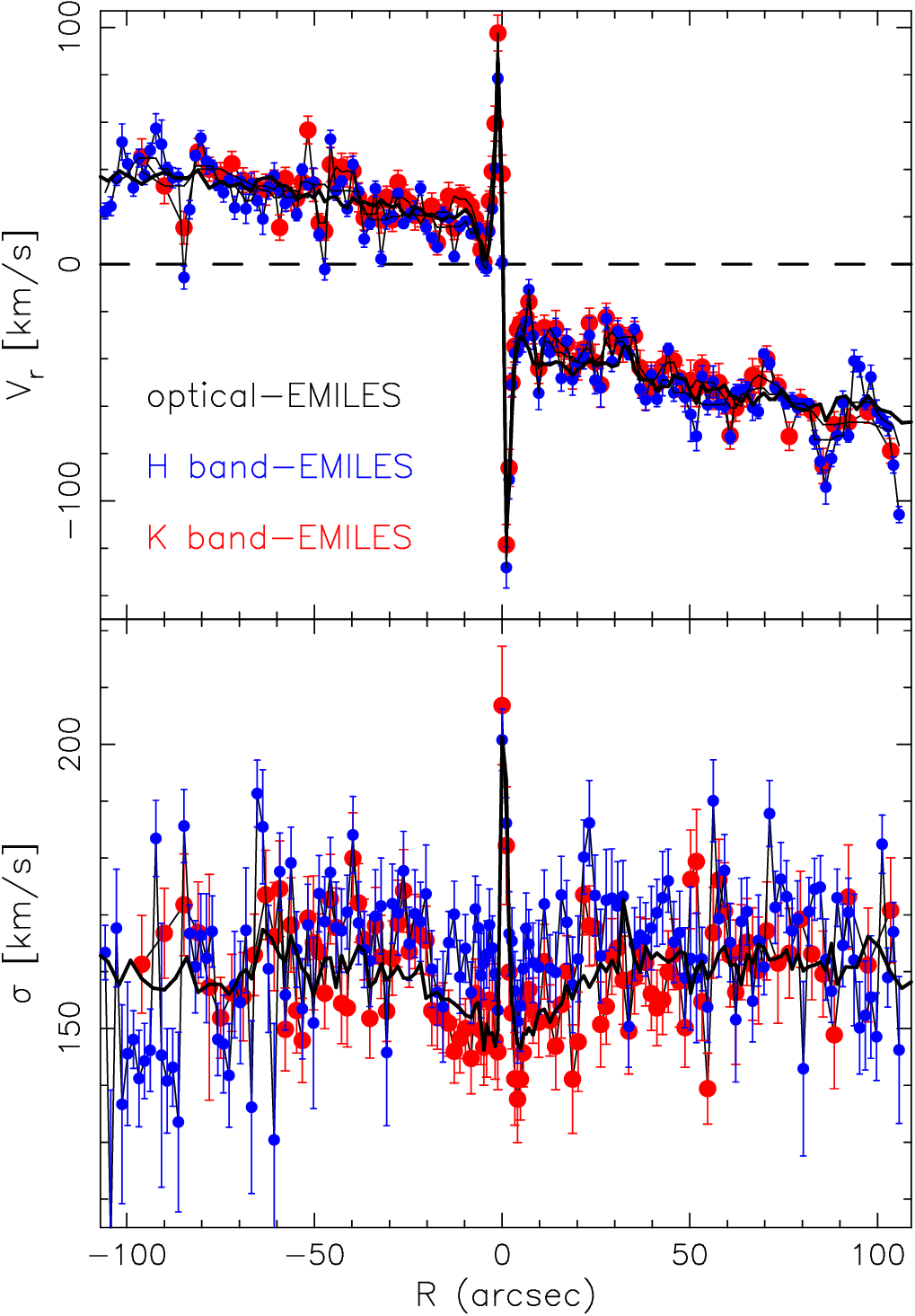}
\end{center}
 \caption{
 Rotation velocity, \VROT , and velocity dispersion, \SIG , for the bulge of M31, as a function of galactocentric distance $\rm R$. Negative and positive values of R correspond to opposite sides of the LUCI slit. The black dashed line in the upper panel marks the value of zero. The black solid curves show the kinematics from the optical (Paper I), while the blue and red curves plot the kinematics obtained with EMILES (Teramo) models in H- and K-band, respectively.  Error bars correspond to 1-sigma uncertainties.
 }
   \label{fig:kin_emiles}
\end{figure}

\begin{figure}
\begin{center}
 \leavevmode
 \includegraphics[width=8cm]{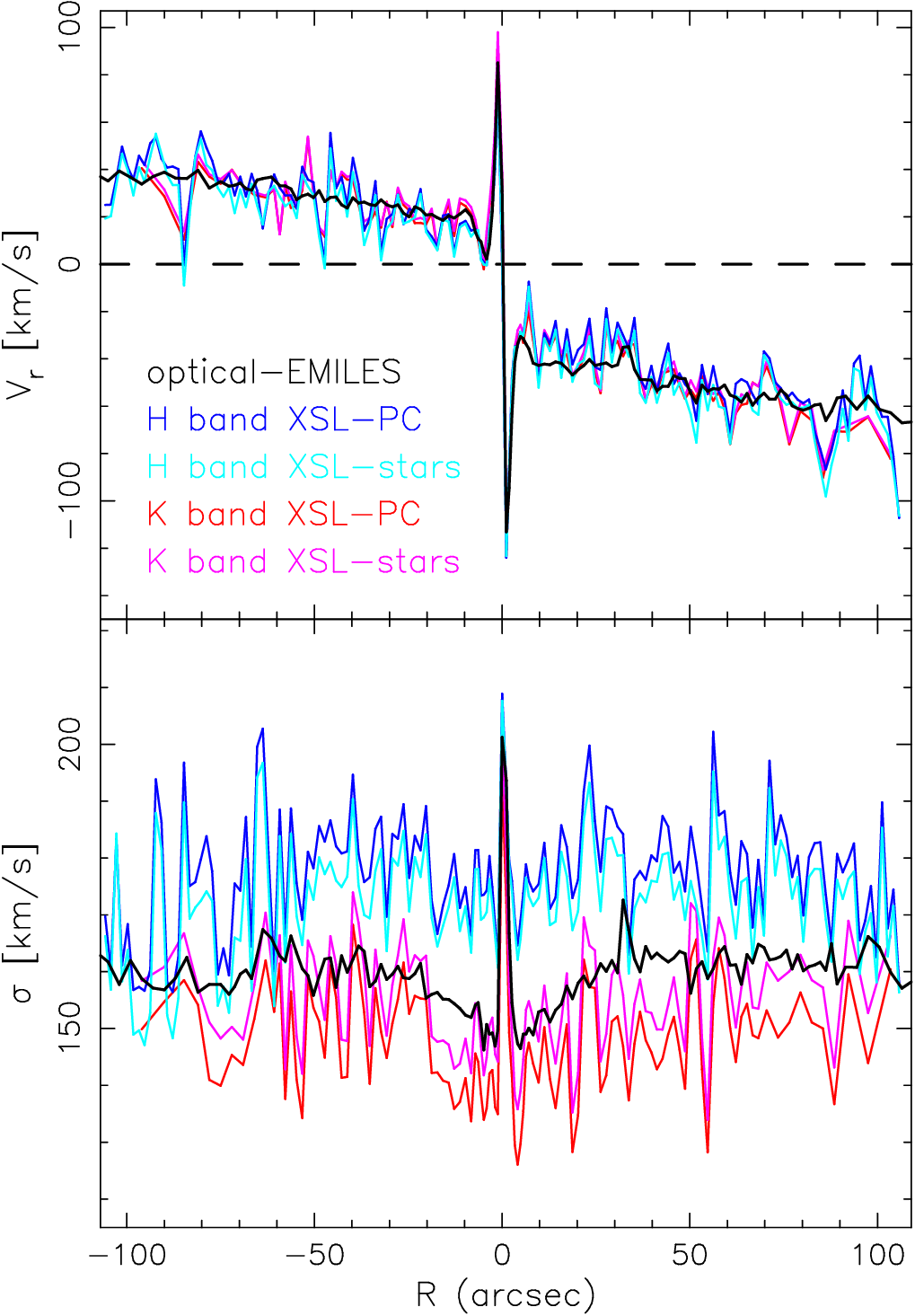}
\end{center}
 \caption{
Same as Fig.~\ref{fig:kin_emiles} but using XSL models to derive the kinematics of the bulge in the H and K band (see the text).  Uncertainties on individual measurements are similar to those shown in Fig.~\ref{fig:kin_emiles}, but are not reported in the Figure in order to make differences among curves clearer.
 }
   \label{fig:kin_xsl}
\end{figure}

We measured the kinematics of the M31 bulge, namely rotation velocity \VROT , and velocity dispersion, $\sigma$, for both the H and K band, using adaptively binned spectra with a minimum $S/N$ ratio of 70~\AA$^{-1}$ (see Sec.~\ref{subsec:binning}), and running the software {\sc pPXF}~\citep{Cap:2004, Capp:17}.
Fig.~\ref{fig:kin_emiles} shows the profiles of \VROT\ and \SIG\ obtained by feeding pPXF with EMILES 1SSP models, based on Teramo models. No significant differences were found when using models based on Padova isochrones. The Figure shows that in both H- and K-band, the kinematics is fully consistent with that derived from the optical spectral range (see the black curves). Fig.~\ref{fig:kin_xsl} shows the results obtained with XSL-Parsec 1SSP models. In this case, while the \VROT\ profiles are consistent with the optical, the \SIG\ profiles are offset with respect to the optical ones, by, on average, $-7$ and $+14$~\kms, respectively.  Using XSL stars, rather than the 1SSP's (see labels in upper panel of the Figure), reduces the offset by only $\sim 1-2$~\kms, showing that the discrepancy is intrinsic to the stellar spectra. At the moment, the origin of this issue is not clear.

\section{Radial trend of C4668 in the bulge}
\label{app:C4668}
\begin{figure}
\begin{center}
 \leavevmode
 \includegraphics[width=8cm]{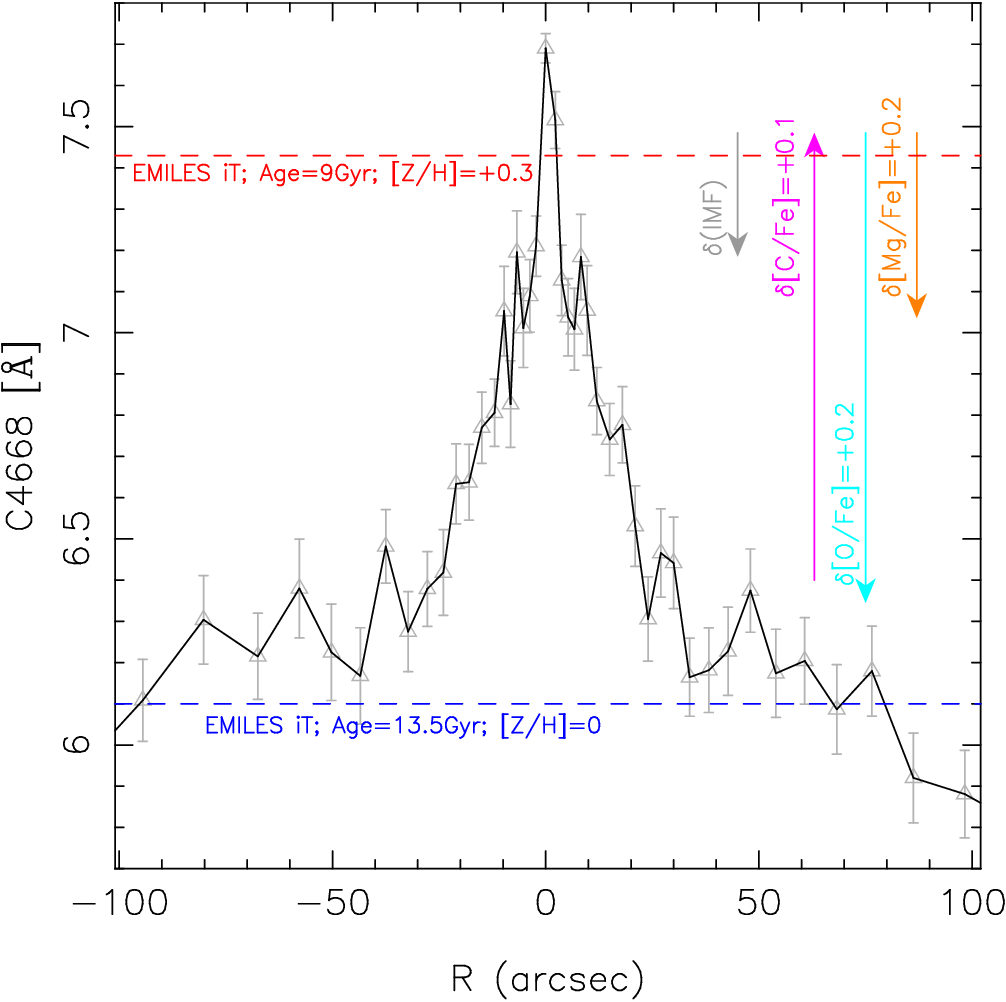}
\end{center}
 \caption{
Radial profile of the C4668 spectral index, from the optical GTC spectroscopy of the M31 bulge (see the text). The profile is plotted over the same radial range as the CO indices (see Fig.~\ref{fig:CO_profiles}). Error bars correspond to 1~sigma uncertainties on spectral indices. The horizontal dashed lines mark the predictions of 1SSP EMILES models, with Teramo isochrones, for the central and outer regions of the bulge (see Fig.~\ref{fig:CO_comp_models}), i.e. for an age of 9~Gyr and $[Z/H]=+0.3$ (red), and for an age of 13.5~Gyr and $[Z/H]=0$ (blue), respectively.  The arrows are the same as those in Fig.~\ref{fig:CO_comp_models}), i.e. the pink, cyan and, orange arrows show the effect of varying \cfe, \ofe, and \mgfe\ abundance ratios, according to CvD18 models, while the gray arrow shows the effect of varying the IMF from \gammab$=1.3$ (Kroupa-like), to \gammab$=2.5$.
 }
   \label{fig:C4668}
\end{figure}

Fig.~\ref{fig:C4668} plots the radial profile of the C4668 index for the bulge of M31, as measured from the optical GTC spectra (see Sec.~\ref{sec:obs}). The C4668 measures molecular C$_2$ absorption, with a strong sensitivity to \cfe , as shown by the pink arrow in the Figure (see~\citealt{JTM:2012, CvD12a}). The index anticorrelates with \ofe\ and \mgfe\ (at fixed total metallicity), as shown by the cyan and orange arrows in the Figure. In contrast to most of the NIR CO indices, showing flat radial trends (see Fig.~\ref{fig:CO_profiles} and Sec.~\ref{sec:COrad}), the C4668  exhibits a steep radial profile, increasing significantly toward the center of the M31 bulge. The steep profile results from a combined effect of metallicity, as the index increases with \zh\ (see horizontal dashed lines in the Figure), as well as the combined effect of \cfe\ and other abundance ratios. In contrast to CO indices, the effect of the IMF is subdominant with respect to that of metallicity and abundance ratios (see the green arrow in the Figure), while in the case of CO lines, the IMF tends to compensate the effect of other stellar populations properties, leading to flat radial trends.

\end{appendix}
\end{document}